\documentclass[12pt,onecolumn,draftclsnofoot]{IEEEtran}   % Xiang: use this in the submitted journal version

\usepackage{cite}
\usepackage{graphicx,color,epsfig,rotating}
\usepackage{amsfonts,amsmath,amssymb,bbm}
\usepackage{setspace}
\usepackage{algorithm}
\usepackage[algo2e]{algorithm2e} 
\usepackage{subcaption}
\usepackage{mdwtab}
\usepackage{placeins}
\usepackage{psfrag, graphicx}
\usepackage[latin1]{inputenc}
\usepackage{multirow}
\usepackage{stfloats}
\usepackage{tabularx} 
\usepackage{booktabs} 
\usepackage{url}
\usepackage{bm}
\usepackage{geometry}
\usepackage{pstricks}
\long\def\comment#1{}

\newtheorem{example}{Example}%[section]
\newtheorem{theorem}{Theorem}%[section]
\newtheorem{lemma}{Lemma}%[section]
\newtheorem{remark}{Remark}%[section]
\newtheorem{definition}{Definition}

\def \rz{{R_Z}}

\let\tbf\textbf
\let\tit\textit

\newcommand{\cds}{conditional disclosure of secrets\xspace}

\newcommand{\ie}{i.e.\xspace}

\newcommand{\Ip}{In particular\xspace}

\newcommand{\comm}{communication\xspace}

\newcommand{\achvb}{achievable\xspace}

\newcommand{\achvblty}{achievability\xspace}

\newfont{\bbb}{msbm10 scaled 700}

\newfont{\bb}{msbm10 scaled 1100}

\newcommand{\Am}{{\bf A}}
\newcommand{\Bm}{{\bf B}}

\newcommand{\Ac}{{\cal A}}
\newcommand{\Bc}{{\cal B}}

\newcommand{\Ic}{{\cal I}}

\newcommand{\Xc}{{\cal X}}
\newcommand{\Yc}{{\cal Y}}

\newcommand{\eqdef}{\stackrel{\Delta}{=}}

\newcommand{\be}{\begin{equation}}
\newcommand{\ee}{\end{equation}}
\newcommand{\bea}{\begin{eqnarray}}
\newcommand{\eea}{\end{eqnarray}}

\SetKwInput{KwData}{Input}
\SetKwInput{KwResult}{Output}
\allowdisplaybreaks 
\graphicspath{{./images/}}

\allowdisplaybreaks

\begin{document}

\title{Graph-Theoretic Characterization of the Noise Capacity of Conditional Disclosure of Secrets}

\author{
Zhou Li,~\IEEEmembership{Member,~IEEE},
Siyan Qin,~\IEEEmembership{Member,~IEEE},
Xiang~Zhang,~\IEEEmembership{Member,~IEEE},
Jihao Fan,~\IEEEmembership{Member,~IEEE},
Haiqiang Chen,~\IEEEmembership{Member,~IEEE},
and Giuseppe Caire,~\IEEEmembership{Fellow,~IEEE}
\thanks{Part of this work \cite{Zhou_Qin_Xiang_CDS} was presented at the 2025 IEEE International Symposium on Information Theory, Ann Arbor, Michigan, USA.
}
\thanks{Z. Li, S. Qin, and H. Chen are with the School of Computer, Electronics and Information, 
Guangxi University, Nanning 530004, China (e-mail: lizhou@gxu.edu.cn, 2413302012@st.gxu.edu.cn, and haiqiang@gxu.edu.cn).}

\thanks{X. Zhang and  G. Caire are with the Department of Electrical Engineering and Computer Science, Technical University of Berlin, 10623 Berlin, Germany (e-mail: \{xiang.zhang, caire\}\@tu-berlin.de).
}

\thanks{J. Fan is with the School of Cyber Science and Engineering, Nanjing University of Science and Technology, Nanjing 210094, China and also with  the Laboratory for Advanced Computing and Intelligence Engineering, Wuxi 214083, China (e-mail: jihao.fan@outlook.com).
}
}

\maketitle

\begin{abstract}
In the problem of conditional disclosure of secrets (CDS), two parties, Alice and Bob,  have an input $x \in \mathcal{X}$ and $y \in \mathcal{Y}$, respectively, and share a common secret. 
Let $f:\Xc \times \Yc \mapsto \{0,1\}$ be a function that maps the input pair $(x,y)$ to a binary output.
Alice and Bob aim to reveal the secret to a third party, Carol, through an error-free channel, as efficiently as possible if $f(x,y)=1$.
In contrast, when $f(x,y)=0$, the secret should not be revealed to  Carol.
To protect the secret, Alice and Bob share a common noise variable that is unknown to Carol. 
This work aims to determine the noise capacity of CDS, which is
defined as the maximum number of secret bits that can be securely revealed to Carol per bit of the noise variable.

We first derive the necessary and sufficient conditions on the function $f$ -- which can be represented by a graph -- for the extremal case where the CDS noise capacity attains its maximum value of \(1\).
Second, we develop novel converse bounds on the noise rate for all linear schemes.
In particular, this bound is equal to  $\frac{(\rho - 1)(d - 1)}{\rho d - 1}$ if $\rho$ is finite, and equal to $ \frac{d-1}{d} $ if $\rho$ is infinite, where 
$\rho$ denotes the covering parameter of the graphical representation of $f$ (referred to as the \emph{CDS graph}) and $d$ denotes the number of unqualified edges (i.e., edges for which $f(x,y)=0$) in the associated unqualified path.
Third, under the maximal communication efficiency constraint, \ie, when the message size is equal to the secret size, we refine the proposed converse bounds based on a careful inspection of the qualified components and their interconnections in the CDS graph.  Moreover, we show the achievability of the proposed converse bounds through a CDS instance with cyclic qualified edges and one unqualified path. 
% The proposed graph-theoretic framework not only deepens the understanding of how structural properties of CDS affect noise efficiency but also establishes a unified approach to analyzing security-efficiency trade-offs in information-theoretic cryptography.
The proposed graph-theoretic framework explicitly links the noise efficiency limits of CDS to two fundamental structural parameters -- the residing unqualified path distance and the covering parameter -- providing a concrete and systematic method for analyzing CDS under arbitrary graph topologies.
\end{abstract}


\section{Introduction}
\label{sec:introduction}

Secure communication systems have long been central to cryptography and information theory, motivating extensive studies on the trade-offs between computational and information-theoretic security. While computational security depends on hardness assumptions, information-theoretic security offers unconditional protection even against unbounded adversaries, inspiring growing interest in its potential for multi-user systems. Classical cryptographic problems, such as secure multi-party computation and secret sharing \cite{cramer_damgard_nielsen_2015, Shamir_SecretSharing}, typically prioritize correctness and security with limited use of Shannon-theoretic tools. This gap has attracted attention from the information theory community, leading to Shannon-theoretic models for secure communication and storage \cite{liang2009information, bloch2011physical, yener2015wireless} and renewed study of problems like private information retrieval \cite{Sun_Jafar_PIR, Banawan_Ulukus, Zhou_Sun_Fu, zhang2021fundamental}, secure distributed storage \cite{Li_Sun_SecureStorage, Li_Sun_Storage, Lee_Abbe, Data_Prabhakaran_Prabhakaran}, and secure computation \cite{Zhao_Sun_SMP, yu2018lagrange, chang2018capacity}, revealing their efficiency and scalability advantages.

\begin{figure}[h]
\begin{center}
\includegraphics[width=.35\textwidth]{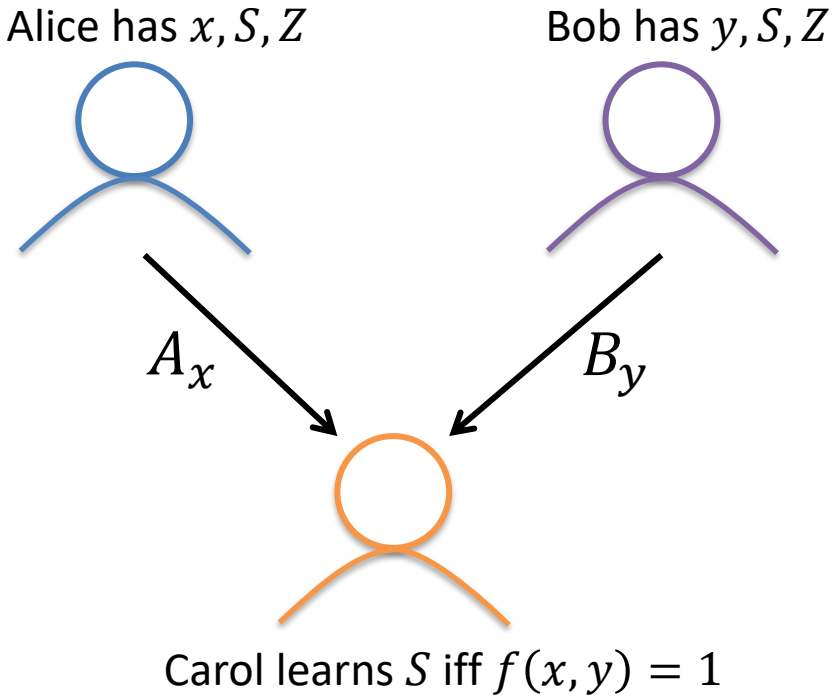}
%\vspace{-0.15in}
\caption{\small  Illustration of the conditional disclosure of secrets (CDS) problem. 
    Alice and Bob have private inputs $x$ and $y$, respectively, i.e., Alice does not know about $y$ and Bob does not know about $x$, and share a secret $S$ along with a common noise variable $Z$ that protects the secret. 
    A publicly known function $f(x, y)$ specifies the disclosure condition: when $f(x, y)=1$, they encode $S$ into their transmitted messages so that Carol can recover it; 
    when $f(x, y)=0$, the messages depend only on $Z$, ensuring that Carol learns nothing about $S$.
}
%\vspace{-0.3in}
\label{figprob}
\end{center}
\end{figure}

The conditional disclosure of secrets (CDS) problem (see Fig.~\ref{figprob}) represents a fundamental challenge in secure multiparty computation. It involves a scenario where two parties, Alice and Bob, hold private inputs and share a common secret, which they aim to reveal to a third party, Carol, under specific conditions determined by their inputs. If the condition is satisfied, Carol should be able to recover the secret with certainty. Conversely, when the condition is not met, no information about the secret should be leaked. This dual objective of correctness and security creates a complex design space for efficient and robust CDS protocols.

CDS has found applications in various real-world cryptographic systems. For instance, in secure voting \cite{Benaloh1994}, a vote tally is revealed only if specific rules are satisfied, ensuring the confidentiality of individual votes. Similarly, in privacy-preserving data aggregation \cite{He2006,Bista2010}, sensitive data is disclosed only under pre-defined conditions, protecting participant privacy. The conditional nature of disclosure makes CDS a cornerstone for privacy-preserving technologies in distributed and resource-constrained environments.

Designing efficient CDS protocols requires balancing resource usage--including communication overhead and randomness consumption--while adhering to strict security guarantees.  Existing studies  have exclusively focused on communication efficiency, aiming to minimize the amount of data exchanged between  different parties. However, noise efficiency, a critical aspect of CDS schemes, has not been sufficiently explored. Common noise, usually implemented as random bits, plays a central role in CDS as  it is required to protect the secret in unqualified conditions. In practical IoT and distributed systems, generating random noise bits consumes extra power of the devices. Therefore, optimizing the noise generation efficiency is a crucial aspect of CDS.

\subsection{Motivation}
\begin{figure}[h]
\begin{center}
\includegraphics[width=.35\textwidth]{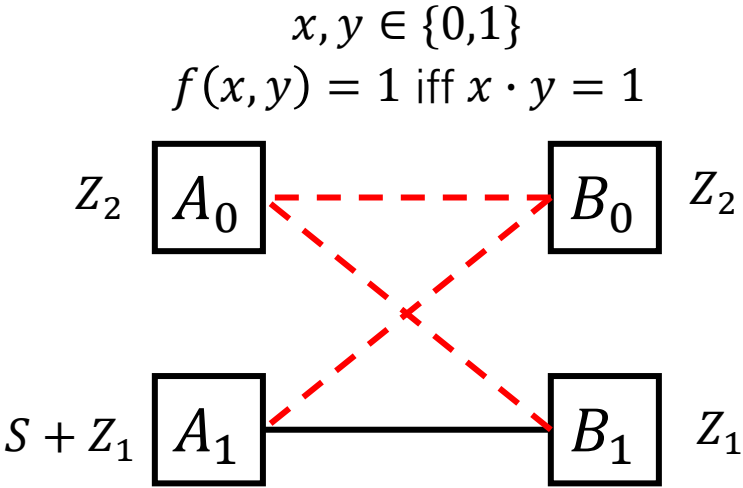}
%\vspace{-0.15in}
\caption{\small The secret is disclosed if and only if $x=y=1$ (i.e., from $A_1, B_1$). 
}
%\vspace{-0.15in}
\label{fig:ex0}
\end{center}
\end{figure}

Although CDS has been extensively studied in the context of communication efficiency, where communication rate refers to the amount of information in bits transmitted per secret revealed or per query executed, the optimization of noise generation efficiency remains underexplored. Earlier work \cite{Li_Sun_CDS}\cite{Li_Sun_linearCDS} introduced the concept of \emph{communication capacity}--defined as the maximum number of secret bits that can be disclosed per bit of total communication--and demonstrated that aligning noise with messages allows CDS schemes to achieve high communication rates while maintaining security. Building on this, investigations into the linear communication capacity of CDS schemes led to upper bounds for linear coding strategies and the identification of structural properties enabling near-optimal performance. These studies established a comprehensive framework for understanding and optimizing communication rates. 
However, noise in these works was primarily treated as a tool to facilitate secure communication, rather than as a metric to be optimized. This perspective ignores the critical role of noise rates, especially in resource-constrained environments such as IoT systems and federated learning, where efficient noise utilization is key to scalability and feasibility.
To address this gap, in this paper, besides the conventional \comm rate, we also focus on optimizing the noise rate.
The relationship between noise usage and secret disclosure is examined, along with the interplay between noise rate and the graphical structure of the CDS problem.
A representative application of CDS can be described as follows. Alice and Bob aim to disclose a secret (e.g., a business plan) to Carol only when both parties decide to collaborate. The Boolean function \( f(x, y) \) specifies the condition under which such mutual agreement occurs. To preserve privacy, neither Alice nor Bob reveals their individual input; although the function \( f \) is public, its evaluation \( f(x, y) \) generally remains unknown to either participant, since Alice observes only \( x \) and Bob only \( y \). The CDS protocol guarantees that Carol receives the secret \emph{if and only if} the collaboration condition is satisfied.  
For example, let Alice's input be \( x \in \{0,1\} \), where \( x=1 \) indicates willingness to collaborate and \( x=0 \) otherwise, and let Bob's input \( y \in \{0,1\} \) be defined similarly. Then \( f(x, y) = xy \), implying that Carol can recover the secret only when \( x = y = 1 \) (see Fig.~\ref{fig:ex0}). 
In this manner, the collaboration is realized in a distributed and secure manner.
This approach provides new insights into the fundamental limits of CDS and paves the way for designing more practical and scalable secure communication systems.
%\vspace{-0.1in}

\subsection{Related Work}

%The \cds (CDS) problem, a fundamental cryptographic primitive, has attracted attention in recent years. 
%CDS involves enabling two parties to disclose a shared secret to a third party only when a  predefined  condition is satisfied. Correctness requires that the third party  should  reliably recover the secret when the predefined  condition on the inputs of the relevant parties are satisfied. Security refers  to the constraint  that  the  third party should infer nothing about the secret if such conditions are not satisfied.
Early studies on CDS focused on minimizing the communication cost of these protocols under computational security assumptions \cite{SymPIR, Gay_Kerenidis_Wee, Applebaum_Arkis_Raykov_Vasudevan}. Recent works have extended these investigations to consider amortized rates for general CDS instances, as in \cite{applebaum2018power}, where the focus is on approximating worst-case rates rather than characterizing exact capacities.
Meanwhile, advances in Shannon-theoretic analysis have introduced new approaches to CDS. Inspired by interference alignment techniques originally developed for interference networks \cite{Jafar_FnT, Jafar_TIM}, the noise and message alignment framework was adapted to the CDS context by Li and Sun \cite{Li_Sun_CDS}. This approach has proven effective for characterizing the linear capacity of specific high-rate CDS instances, providing both converse and achievability results. However, the general linear capacity of CDS remains  open, with many instances yet to be fully understood.
Beyond CDS, related work has explored broader themes in secure communication, such as anonymous communication \cite{Sun_Anonymous}, secure aggregation in distributed networks \cite{so2022lightsecagg,so2021turbo}, and the use of algebraic coding techniques for improving efficiency \cite{yu2018lagrange, chang2018capacity}. These studies demonstrate the versatility of information-theoretic methods in addressing a wide range of cryptographic challenges, underscoring the value of applying these tools to foundational problems like CDS.
The present work builds on these developments by focusing specifically on the linear capacity of CDS. While prior studies such as \cite{Li_Sun_CDS} have provided insights into high-capacity scenarios, this paper aims to expand the scope by identifying general linear converse bounds and developing systematic approaches to linear scheme design. Through this lens, we seek to bridge the gap between cryptographic formulations of CDS and their Shannon-theoretic counterparts, advancing the understanding of CDS in both theory and practice.

\subsection{Summary of Contributions}
This paper advances the understanding of CDS by providing a comprehensive and rigorous framework for optimizing noise efficiency, addressing critical gaps in existing research. Specifically, based on a graph-theoretical framework for CDS, this work makes the following contributions:

\begin{itemize}
    \item We introduce a graphical framework that transforms a CDS instance into a graph capturing conditional disclosure constraints. Based on this framework, we derive upper bounds on noise capacity and reveal how noise can be optimally allocated. We establish the necessary and sufficient conditions for achieving the maximum noise capacity of 1, offering a foundational understanding of how optimal noise utilization can be realized in CDS schemes (see Theorem \ref{thm:1noiserate}). 

    \item We derive a general upper bound on the linear noise rate, defined as the maximum noise rate achievable using linear CDS schemes, in cases where the noise capacity exceeds 1. \Ip, this bound is equal to $\frac{(\rho-1)(d-1)}{\rho d-1}$ when $\rho$ is finite, and equal to $(d-1)/d$ when $\rho$ is infinite, where 
            $\rho$ denotes the covering parameter of the graphical representation of $f$ (referred to as the \emph{CDS graph}) and $d$ denotes the number of unqualified edges in the associated unqualified path (see Theorem \ref{thm:linearnoiserate1}).

    \item Under the maximal communication efficiency constraint, \ie, when the message size is equal to the secret size, we refine the upper bound for the linear noise rate, offering deeper insights into the constraints imposed by this relationship (see Theorem \ref{thm:linearnoiserate2}). This refinement highlights the impact of structural relationships between message and secret sizes on the performance of linear coding strategies.

    \item   Finally, we show the achievability of the proposed converse bounds through an CDS instance with cyclic qualified edges and one unqualified path (see Theorem \ref{thm:ach}). 
\end{itemize}

\emph{Notation}. Throughout the paper, the following notations are used. 
$[m:n]\eqdef \{m, m+1, \cdot, n\}$ if $m \le  n$ and $[m:n]= \emptyset$ if $m<n$. We write $[1:n]$ as $[n]=\{1,\cdots, n\}$ for brevity. 
Bold capital letters $\Am, \Bm \cdots$ represent matrices, and calligraphic letters $\Ac, \Bc\cdots$ represent sets. 
$\Ac  \times \Bc \eqdef \{(x,y):x \in \Ac, y\in  \Bc\} $ denotes the Cartesian product of $ \Ac$ and $\Bc$. Also define $ \Ac \backslash \Bc \eqdef  \{x\in \Ac: x\notin \Bc\} $.

\emph{Paper Organization.} The remainder of the paper is organized as follows.
Section~\ref{sec:model} formulates the CDS problem and introduces the relevant graph-theoretic definitions.
Section~\ref{sec:main results} summarizes the main theoretical results and presents illustrative examples.
Sections~\ref{pfthm1}--\ref{pfthm4} provide detailed proofs of Theorems~\ref{thm:1noiserate}--\ref{thm:ach}, respectively, together with discussions on achievability and tightness.
Finally, Section~\ref{conclus} concludes the paper and outlines possible directions for future work.

\section{Problem Statement}\label{sec:model}
The \cds (CDS) problem involves three parties--Alice, Bob, and Carol. 
Let $(x, y)$ be a pair of \emph{inputs} from the set $\mathcal{I} \subseteq \{1,2,\cdots, X\} \times \{1,2,\cdots, Y\}$. Alice has access only to $x$, while Bob has access only to $y$. Alice and Bob share a secret $S$, which consists 
of $L$ i.i.d. uniform symbols from some finite field $\mathbb{F}_p$. Alice and Bob share a common noise variable $Z$, which is independent of the secret $S$ and consists of $L_Z$ i.i.d. uniform symbols from $\mathbb{F}_p$, introduced to conceal information about $S$ when the security constraints cannot be met using $S$ alone.
\begin{eqnarray}
H(S) = L, ~H(Z) = L_Z, ~H(S, Z) = H(S) + H(Z) = L + L_Z. \label{sz_ind}
\end{eqnarray}
Note that the above  entropy terms  are  in $p$-ary units.

Alice and Bob aim to share the secret $S$ with Carol only if $f(x,y)=1$, where $f$ is a publicly known binary function defined over  the input domain $\mathcal{I}$. If $f(x,y)=0$, Carol should not gain any information about $S$. To achieve this, Alice transmits a message $A_x$, while Bob transmits $B_y$, both of which are derived from the secret $S$ and the noise variable $Z$ which are hidden from Carol:
\begin{eqnarray}
H\left(A_x, B_y | S, Z\right) = 0, \;\forall (x,y) \in \Ic   \label{det} 
\end{eqnarray}
For simplicity, we restrict attention to the case where each message consists of $N$ independent and uniformly distributed symbols over $\mathbb{F}_p$, so that %\textcolor{red}{(This condition is not necessary in general; however, we impose this restriction in order to derive the upper bound in Theorem 2. Without this restriction, the problem becomes technically nontrivial.)}
\begin{eqnarray}
   H(A_x)=H(B_y)=N. \label{messagezise}
\end{eqnarray}

%\xz{Let the \msgs $A_x$ and $B_y$ contain at most $L_A$ and $L_B$ symbols from some finite field, \resp, \ie, $L_A \eqdef \max_{x\in \Xc}|A_x| , L_B \eqdef \max_{y\in \Yc}|B_y|    $. It holds that $L_A\ge \max_{x\in \Xc}H(A_x), L_B\ge \max_{y\in \Xc}H(B_y)$.   }

Carol does not know $x$ or $y$, but the protocol is designed so that Carol can recover the secret $S$ from $A_x$ and $B_y$ if $f(x,y)=1$. Otherwise, if $f(x,y)=0$, the pair $(A_x,B_y)$ must remain independent of $S$, ensuring that no information about $S$ is revealed to Carol.
For any $(x,y) \in \mathcal{I}$, the following correctness and security constraints should be satisfied:
\begin{eqnarray}
&& [\mbox{Correctness}] ~~H(S | A_x, B_y) = 0,  ~~~~~~~~~\mbox{if}~f(x,y) = 1. \label{dec} \\
&& [\mbox{Security}]~~~~~~H(S | A_x, B_y) = H(S), ~~~~\mbox{if}~f(x,y) = 0. \label{sec}
\end{eqnarray}
The collection of the mappings from $\{x,y,S,Z\}$ to the messages $A_x, B_y$ 
is called a CDS scheme.

In our previous work \cite{Li_Sun_CDS, Li_Sun_linearCDS},
the \comm rate--defined  as $R=L/(2N)$--was studied as the primary objective to optimize. However, the randomness consumption aspect, represented by the efficiency of the noise usage, has not been investigated. To address, 
in this work,
we focus on the \emph{noise rate} $R_Z$  of the CDS problem. In particular, the noise rate $R_Z$ represents how many symbols of the secret that  can be securely disclosed per symbol of noise variable $Z$, \ie, 
\begin{eqnarray}
R_Z = \frac{L}{L_Z}. 
\end{eqnarray}
A noise rate $R_Z$ is said to be achievable if there exists a CDS scheme which simultaneously satisfy  the correctness constraint (\ref{dec}) and the security constraint (\ref{sec}). 
The \emph{capacity} of the CDS problem, denoted by $C$, is defined as the supremum of all achievable noise rates.

\subsection{Graph-Related Definitions}
To present our results, we use several graph-theoretic concepts related to $G_f = (V, E)$, where $V$ and $E$ denote the sets of nodes and edges, respectively. In our setting, the graph $G_f$ is \emph{inherently bipartite}: one part consists of the messages sent by Alice and the other consists of the messages sent by Bob,
$V = \{A_1,\ldots,A_X\} \cup \{B_1,\ldots,B_Y\}.$
Edges exist only between nodes of different types, and an unordered pair $\{A_x,B_y\}$ is included in $E$ precisely when $(x,y)\in\mathcal{I}$.
Since each message corresponds to a unique node, the terms ``message'' and ``node'' are used interchangeably throughout the paper. Each edge $\{A_x,B_y\}\in E$ is labeled as \emph{qualified} if $f(x,y)=1$, in which case it is drawn as a solid black line, or as \emph{unqualified} if $f(x,y)=0$, in which case it is drawn as a dashed red line. 
This labeled bipartite graph representation is illustrated in Fig.~\ref{fig1}.

% \textcolor{blue}{In the graph $G_f$, the set of nodes represents all the messages transmitted by Alice and Bob, namely
% $V = \{A_1, \ldots, A_X, B_1, \ldots, B_Y\}$.
% Because each message corresponds uniquely to a node, the two terms message and node are used interchangeably in this paper.
% The edge set $E$ is composed of unordered pairs $\{A_x, B_y\}$ satisfying $(x, y) \in \mathcal{I}$.
% We assign each edge a type through a mapping $t: E \rightarrow {0,1}$.
% When $f(x,y) = 1$ (equivalently, $t(A_x, B_y) = 1$), the edge $\{A_x, B_y\}$ is referred to as a {\em qualified edge} and depicted as a solid black line.
% Conversely, when $f(x,y) = 0$ (equivalently, $t(A_x, B_y) = 0$), it is referred to as an {\em unqualified edge} and shown as a dashed red line.
% An illustrative example is provided in Fig.~\ref{fig1}.}

%Without loss of generality, we assume that for any node $v\in V$, there exists at least one node $u\in V$ such that $\{u,v\}\in E$ is an unqualified edge. 
%If no unqualified edge is present, meaning that node $v$ is connected solely by qualified edges, $v$ can be designated as the secret $S$. After designating $v$ as the secret, the node and all its associated edges are excluded from further analysis.

\begin{definition}[Qualified/Unqualified Path and Component]
A {\em qualified (unqualified) path} is defined as a sequence of distinct and connected qualified (unqualified) edges. A {\em qualified (unqualified) connected component} refers to a maximal induced subgraph of $G_f$ in which any two nodes are connected by a qualified (unqualified) path.
\end{definition}

For example, in Fig.~\ref{fig2}, the path $P = \{\{A_1, B_1\}, \{B_1, A_2\}, \{A_2, B_2\}\}$ is both a qualified path and a qualified component. Similarly, the path $P = \{\{A_2, B_3\}, \{B_3, A_1\}, \{A_1, B_2\}\}$ is both an unqualified path and an unqualified component.

\begin{definition}[Internal Qualified Edge and Residing Unqualified Path]
A {\em qualified edge} that connects two nodes, denoted as $A_i$ and $B_j$, in an unqualified path is called an {\em internal qualified edge}. The unqualified path with end nodes $A_i$ and $B_j$ is referred to as the {\em residing unqualified path} of the internal qualified edge $\{A_i, B_j\}$.
\end{definition}

For example, in Fig.~\ref{fig2}, consider the unqualified path $P = \{\{A_2, B_3\}, \{B_3, A_1\}, \{A_1, B_2\}\}$. The nodes $A_2$ and $B_2$ are connected by the qualified edge $\{A_2, B_2\}$, which is an internal qualified edge. The unqualified path $P$ is the residing unqualified path of $\{A_2, B_2\}$.

\begin{definition}[Residing Unqualified Path Distance] \label{def:rupd}
For an internal qualified edge $e$ and its residing unqualified path $P$, the number of edges in $P$ is called the {\em residing unqualified path distance} and is denoted as $d(e, P)$. If no residing unqualified path exists, $d(e, P)$ is defined as $+\infty$. Furthermore, $d \triangleq \min_{e, P} d(e, P)$.
\end{definition}

For example, in Fig.~\ref{fig2}, the residing unqualified path distance is $d=3$, corresponding to the edges $
\{A_2, B_3\}$, $\{B_3, A_1\}$, and $\{A_1, B_2\}$.

\begin{definition}[Connected Edge Cover]\label{def:cec}
Consider an internal qualified edge $e$ and a residing unqualified path $P$, with the set of nodes in $P$ denoted as $V_P \subset V$. A {\em connected edge cover} of $V_P$ is a set of connected\footnote{That is, any two nodes in $M$ are connected by a qualified path.} qualified edges $M \subset E$ such that each node in $V_P$ is covered by at least one edge in $M$, and $e \in M$. The size of the connected edge cover for $(e, P)$ is the number of edges in $M$ and is denoted as $\rho(e, P)$. If no such $M$ exists, then $\rho(e, P)$ is defined as $+\infty$. Furthermore, $\rho \triangleq \min_{e, P} \rho(e, P)$.
\end{definition}

For example, in Fig.~\ref{fig3}, consider the internal qualified edge $e = \{A_2, B_2\}$ in the unqualified path $P = \{\{A_2, B_3\}, \{B_3, A_1\}, \{A_1, B_2\}\}$. The nodes in $P$ are $V_P = \{A_2, B_3, A_1, B_2\}$. A connected edge cover of $V_P$ is given by $M = \{\{A_1, B_1\}, \{B_1, A_2\}, \{A_2, B_2\}, \{B_2, A_3\}, \{A_3, B_3\}\}$. In this case, $\rho(e, P) = 5$, as $M$ contains 5 qualified edges. Furthermore, we verify that the minimum value of $\rho(e, P)$ across all internal qualified edges and their associated unqualified path pairs $(e, P)$ is $\rho = 5$.

\begin{definition}[Components of Residing Unqualified Path] \label{dec:crup}
Consider an internal qualified edge $e$ and a residing unqualified path $P$, components of the residing unqualified path is defined as the number of qualified components that are connected to at least one node in $P$. This value is denoted as $Q(e, P)$. Note that the internal qualified edge should be in the same qualified component because these two nodes are in the same qualified edge. If there is no internal qualified edge, then $Q(e, P)$ is defined as $+\infty$. Further, $Q \triangleq \min_{e, P} Q(e, P)$.
\end{definition}

For example, in Fig.~\ref{fig2}, consider the unqualified path $P = \left\{{A_2, B_3}, {B_3, A_1}, {A_1, B_2}\right\}$. The nodes $A_1$, $A_2$, and $B_2$ belong to the same qualified component, whereas $B_3$ belongs to a different qualified component. Thus, $Q = 2$.

\subsection{Linear Feasibility}\label{sec:linear}
In this section, we characterize the feasibility condition of a linear CDS scheme.

\bigskip
\noindent {\bf Linear Scheme:} {For a feasible linear CDS scheme, each message $v$ is a linear function of the secret $S\in \mathbb{F}_p^{L\times 1}$ and the noise $Z\in \mathbb{F}_p^{L_Z\times 1}$. All secret and noise symbols are assumed to be i.i.d.  uniform. We have
\begin{eqnarray}
v = {\bf F}_v S + {\bf H}_v Z, ~{\bf F}_v \in \mathbb{F}_p^{N \times L}, {\bf H}_v \in \mathbb{F}_p^{N \times L_Z} \label{eq:linear}
\end{eqnarray}

Each node $v$ is assumed to connect to at least one unqualified edge\footnote{Consider any node $v$ that is incident only to qualified edges; equivalently, $v$ is not subject to any security constraint. For such a node, we may simply set its message to be the secret $S$ and remove $v$ from the graph. By iteratively applying this procedure to all such nodes, we obtain a reduced graph in which every remaining node is incident to at least one unqualified edge.
}, ensuring that \( I(v; S) = 0 \). Under this assumption, any rows of \(\mathbf{H}_v\) in \(v\) must also remain linearly independent. Since each message \(v\) consists of \(N\) symbols, \(\mathbf{H}_v\) must have a row rank of \(N\). This can be expressed as:
\begin{equation}
    \text{rank}(\mathbf{H}_v) =  N.
\end{equation}

For any edge $\{v, u\}$, consider the overlap between the noise spaces of $v$ and $u$, specifically the intersection of the row spaces of ${\bf H}_v$ and ${\bf H}_u$. Let ${\bf P}_v$ and ${\bf P}_u$ be projection matrices such that:
\begin{eqnarray}
{\bf P}_v {\bf H}_v &=& {\bf P}_u {\bf H}_u, \notag\\
\mbox{rank}({\bf P}_v) &=& \mbox{rank}({\bf P}_u) = \mbox{dim}(\mbox{rowspan}({\bf H}_v) \cap \mbox{rowspan}({\bf H}_u)). \label{eq:proj}
\end{eqnarray}
Intuitively, these projection matrices extract the shared noise components between the two nodes; if the two noise spaces have no overlap, then the intersection is empty and both ${\bf P}_v$ and ${\bf P}_u$ educe to zero matrices, meaning that no shared noise component exists to extract.

For any edge $\{v,u\}$, consider the overlap between the noise spaces of $v$ and $u$, i.e., the intersection of the row spaces of ${\bf H}_v$ and ${\bf H}_u$. Let ${\bf P}_v$ and ${\bf P}_u$ be projection matrices onto this intersection, satisfying
\begin{equation}
{\bf P}_v {\bf H}_v = {\bf P}_u {\bf H}_u, 
\qquad
\operatorname{rank}({\bf P}_v)=\operatorname{rank}({\bf P}_u)
= \dim\big(\operatorname{rowspan}({\bf H}_v)\cap\operatorname{rowspan}({\bf H}_u)\big).
\label{eq:proj}
\end{equation}
Intuitively, ${\bf P}_v$ and ${\bf P}_u$ extract the shared noise components that appear in both vertices.  
If the intersection $\operatorname{rowspan}({\bf H}_v)\cap\operatorname{rowspan}({\bf H}_u)$ is empty, then both ${\bf P}_v$ and ${\bf P}_u$ reduce to all-zero matrices, meaning that no common noise component exists to extract.

\paragraph{Noise cancellation for qualified edges.}
To recover the secret along a qualified edge, the receiver (say, Carol) can perform the noise-cancellation operation
\begin{align}
{\bf P}_v {v} - {\bf P}_u {u}
=
\big({\bf P}_v {\bf F}_v {S} + {\bf P}_v {\bf H}_v { Z}\big)
-
\big({\bf P}_u {\bf F}_u {S} + {\bf P}_u {\bf H}_u { Z}\big) =
\big({\bf P}_v {\bf F}_v - {\bf P}_u {\bf F}_u\big){ S}. 
\end{align}
Here ${S}$ is the secret vector and ${Z}$ is the noise vector.  
The noise term cancels exactly because ${\bf P}_v {\bf H}_v = {\bf P}_u {\bf H}_u$.  
Hence the secret ${S}$ can be retrieved if and only if
\begin{equation}
\operatorname{rank}({\bf P}_v {\bf F}_v - {\bf P}_u {\bf F}_u) \ge L, \label{eq:dec} 
\end{equation}
ensuring that enough independent linear combinations of the secret remain after noise cancellation. This corresponds to the correctness condition.

\paragraph{Security for unqualified edges.}
If the edge $\{v,u\}$ is unqualified, the security condition
\begin{equation}
{\bf P}_v {\bf F}_v = {\bf P}_u {\bf F}_u \label{eq:sec}
\end{equation}
ensures that the secret is also canceled by the same operation:
\begin{equation}
{\bf P}_v {v} - {\bf P}_u {u}
=
\big({\bf P}_v {\bf F}_v - {\bf P}_u {\bf F}_u\big){ S}
= 0.
\end{equation}
Thus no information about the secret ${ S}$ is revealed through the shared noise subspace.

% Then the secret spaces satisfy the following conditions:
% \begin{eqnarray}
% && [\mbox{Correctness}] ~~~ \mbox{rank}({\bf P}_v {\bf F}_v - {\bf P}_u {\bf F}_u) \geq L, ~~~ \text{if $\{u,v\}$ is qualified}; \label{eq:dec} \\
% && [\mbox{Security}] ~~~~~~~~ {\bf P}_v {\bf F}_v = {\bf P}_u {\bf F}_u, ~~~~~~~~~~~~~~~~ \text{if $\{u,v\}$ is unqualified}. \label{eq:sec}
% \end{eqnarray}
% \footnotetext{The term "security" implies that no additional information about the secret is revealed if the edge $\{u,v\}$ is unqualified.}

Next, to streamline future references we abstract two necessary structural properties of any feasible linear scheme. We emphasize that the Message Alignment property (\ref{eq:message}) is exactly equivalent to the security condition (\ref{eq:sec}), while the Noise Alignment property (\ref{eq:noise}) is necessary for the correctness condition (\ref{eq:dec}) but not sufficient without further constraints on the secret-space matrices. Comprehensive proofs are provided in Lemma 6 and Lemma 7 of \cite{Li_Sun_CDS}, with detailed explanations in Section II.B of \cite{Li_Sun_linearCDS}.

\begin{lemma}\label{lemma:align}
\tit{For any linear scheme as defined above and any edge $\{v,u\}$, the following properties hold:}
\begin{eqnarray}
&& [\mbox{Noise Alignment}] ~~
\mbox{dim}(\mbox{rowspan}({\bf H}_v) \cap \mbox{rowspan}({\bf H}_u)) \geq L,  ~~\mbox{if $\{u,v\}$ is qualified}; \label{eq:noise} \\
&& [\mbox{Message Alignment}] ~~~~~~~~~~~~~{\bf P}_v {\bf F}_v = {\bf P}_u {\bf F}_u, ~~~~~~~~~~~~~~~~~~~\mbox{if $\{u,v\}$ is unqualified}. \label{eq:message}
\end{eqnarray}
\end{lemma}

\section{Main Results}
\label{sec:main results}
Our first main result is a  necessary and sufficient condition for all CDS instances such that the noise capacity is $1$ (highest), as stated in Theorem \ref{thm:1noiserate}.

\begin{theorem}\label{thm:1noiserate}
\tit{The noise capacity of CDS is $1$ if and only if there is no internal qualified edge in an unqualified path.}
\end{theorem}

The proof of Theorem \ref{thm:1noiserate} is detailed in Section \ref{pfthm1}. To provide an intuitive understanding, we present two examples. In the first example, the noise capacity condition for $1$ is satisfied, demonstrating that a noise rate of $1$ is achievable.

\begin{figure}[h]
\begin{center}
\includegraphics[width= 2.5 in]{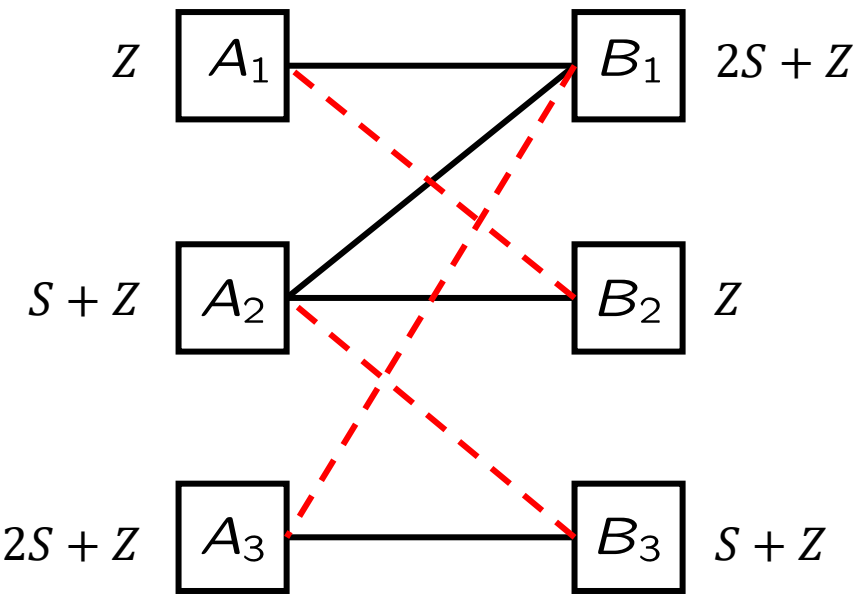}
\caption{\small A CDS instance with the coding scheme that achieves noise rate $1$. 
}
\label{fig1}
\end{center}
\end{figure}

\begin{example}[Achievability of $\rz=1$]
\label{ex1}
Consider the CDS instance depicted in Fig.~\ref{fig1}, represented by the graph $G_f$. It contains three unqualified paths ($\{A_1,B_2\}, \{A_2,B_3\}, \{A_3,B_1\}$), none of which includes any internal qualified edges. It is worth noting that an unqualified edge is treated as an unqualified path or component. As a result, the noise capacity condition for $1$ in Theorem \ref{thm:1noiserate} is satisfied, and Fig.~\ref{fig1} illustrates that a noise rate of $1$ is \achvb.

In this scheme, achieving a noise rate of $1$ requires that every node employ the same noise variable. For example, as illustrated by the code in Fig.~\ref{fig1}, all nodes in the graph $G_f$ utilize the same noise $Z$. For the same unqualified component, all nodes are assigned the same message.
For distinct unqualified components, each node within an unqualified component is assigned a linearly independent combination of the secret and the noise. For example, the three unqualified components are assigned $A_1=B_2=Z$, $A_2=B_3=S+Z$, and $A_3=B_1=2S+Z$, respectively.

The noise capacity is $1$ because the secret consists of $1$ symbol, and $1$ symbol of noise is used. Next, we demonstrate that this scheme satisfies both security and correctness.

\tbf{Security.}
Consider the security of the scheme. Any unqualified edge within the same unqualified component, as well as the nodes in that component, are assigned the same message, ensuring that no information is leaked. Therefore, security is guaranteed. For example, in Fig.~\ref{fig1}, the unqualified edge $\{A_2, B_3\}$ belongs to the same unqualified component, meaning that the nodes $A_2$ and $B_3$ are assigned the same message, $A_2=B_3=S+Z$.

\tbf{Correctness.}
Consider the correctness of the scheme. Any two nodes in a qualified edge belong to different unqualified components, and each component is assigned a linearly independent combination of the secret and noise, allowing the secret to be successfully recovered. Note that there are no internal qualified edges, so any two nodes in the same qualified edge must belong to different unqualified components. For example, in Fig.~\ref{fig1}, the qualified edge $\{A_2, B_1\}$ belongs to the same qualified component, with nodes $A_2$ and $B_1$ belonging to different unqualified components, $(A_2, B_3)$ and $(A_3, B_1)$, respectively. The nodes $A_2$ and $B_1$ are assigned linearly independent combinations of the secret and noise, i.e., $A_2 = S + Z$ and $B_1 = 2S + Z$, from which the secret $S$ can be recovered.
\hfill $\lozenge$
\end{example}

For the second example, the condition in Theorem \ref{thm:1noiserate} is violated such that noise rate $1$ is not achievable. We consider the CDS instance in Fig.~\ref{fig2} as the second example.

\begin{figure}[h]
\begin{center}
\includegraphics[width= 3 in]{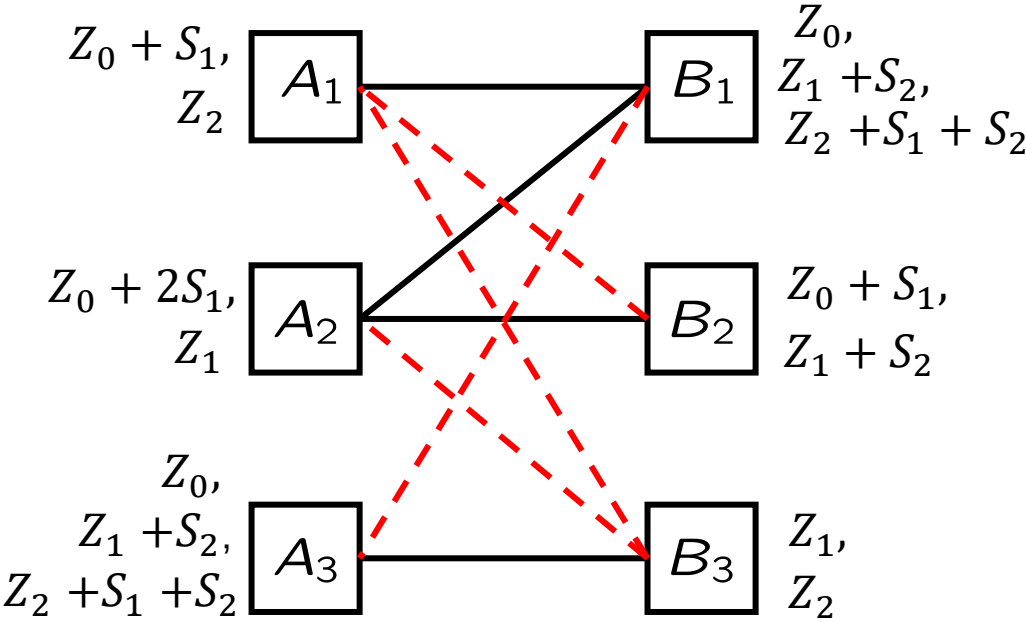}

\caption{\small A CDS instance that has an internal qualified edge $\{B_2, A_2\}$ in a residing unqualified path $(B_2, A_1, B_3, A_2)$. The residing unqualified path distance is $d=3$, i.e., $\{A_2,B_3\},\{B_3,A_1\},\{A_1,B_2\}$. The secret consists of $L = 2$ symbols, $S_1$ and $S_2$, while the noise consists of $L_Z = 3$ independent uniform symbols, $Z_0$, $Z_1$, and $Z_2$. The achieved noise rate is $R = L / L_Z = 2 / 3$.
The scheme achieves this rate by allowing non-uniform message sizes across nodes and therefore does not satisfy the constraint that all nodes have message size $N$. 
}
\label{fig2}
\end{center}
\end{figure}

\begin{example}[Counterexample with Violation]
\label{ex2}
Consider the CDS instance in Fig.~\ref{fig2}. The unqualified path $(B_2, A_1, B_3, A_2)$ contains an internal qualified edge $\{B_2, A_2\}$, violating the noise capacity condition for $R_Z = 1$ in Theorem \ref{thm:1noiserate}, making a noise rate of 1 unachievable. An intuitive explanation by contradiction is as follows.

Suppose the noise rate of 1 is achievable. Then, the size of each message $A_x$ and $B_y$ connected to a qualified edge must be $L_Z = L$ symbols, and the noise in the message must also have size $L_Z = L$ symbols (see Lemma \ref{lemma:size} in Section \ref{sec:thm11}).

For the security constraint, all nodes in the graph $G_f$ must have the same noise variable of size $L_Z = L$ symbols (see Lemma \ref{lemma:noise1} in Section \ref{sec:thm11}). For example, in Fig.~\ref{fig2}, $A_1$, $B_1$, $A_2$, $B_2$, $A_3$, and $B_3$ must use the same noise.

Next, consider any unqualified edge. Given that the noise space fully overlaps, the message space must also fully overlap to prevent leaking information about the secret (see Lemma \ref{lemma:message1}). For example, $B_2$ must equal $A_1$ in Fig.~\ref{fig2}. Then, by sub-modularity, for any unqualified path, the message spaces must fully overlap (see Lemma \ref{lemma:message2}). For example, in Fig.~\ref{fig2}, we must have $B_2 = A_1 = B_3 = A_2$ for the unqualified path $(B_2, A_1, B_3, A_2)$.

Finally, the presence of an internal qualified edge $\{B_2, A_2\}$ leads to a contradiction. On one hand, since $A_2$ and $B_2$ lie on the same unqualified path, they must use the same message. On the other hand, as they are connected by a qualified edge, $A_2$ must be linearly independent of $B_2$.
So the edge ${B_2, A_2}$ cannot be qualified.
\hfill $\lozenge$
\end{example}

Note that noise rate $1$ is the highest for any graph $G_f$ such that each node $v$ has at least one unqualified edge and the noise size cannot be smaller than the secret size, i.e., $L_Z \geq L$ and $R = L/L_Z \leq 1$. As the noise capacity for $R_Z=1$ condition is fully characterized, we proceed to scenarios where noise rate $1$ is not achievable. Our second main result yields an upper bound for the linear noise capacity for any CDS, stated in Theorem \ref{thm:linearnoiserate1}.

%We find the linear noise capacity upper bound for all CDS instances.
%Our second main result is the linear noise capacity characterization of all CDS instances, stated in Theorem \ref{thm:linearnoiserate1}.

\begin{theorem}\label{thm:linearnoiserate1}
\tit{For any CDS problem instance, the linear noise rate of any linear coding scheme is upper bounded by}
\begin{eqnarray}
    R^{(\mbox{\scriptsize linear})}_Z \leq \left\{
    \begin{array}{cl}
        \frac{(\rho-1)(d-1)}{\rho d-1}& ~\rho < +\infty\\
         \frac{d-1}{d}& ~\rho= +\infty
    \end{array}
    \right.
\end{eqnarray}
\end{theorem}

\begin{remark}
When $\rho = +\infty$, for any internal qualified edge $e$, no set of connected edges exists that can cover all nodes in the unqualified path containing $e$ (see Definition \ref{def:cec}). This is equivalent to say that there is no internal qualified edge within any qualified component, which reduces to the linear noise rate upper bound $(d-1)/d$.
\end{remark}

\begin{remark}
When $\rho=+\infty$, and $d = +\infty$, we have that there is no residing unqualified path connected to the internal qualified edge, i.e., there is no internal qualified edge, which reduces to the linear noise capacity $R_Z=1$ condition in theorem 1.
\end{remark}

The proof of theorem 2 is provided in Section \ref{pfthm2}. Similar to Theorem 1, we present two examples to offer an intuitive understanding. In the first example, a connected edge cover exists, i.e., $\rho < +\infty$.

\begin{figure}[h]
\begin{center}
\includegraphics[width=.6\textwidth]{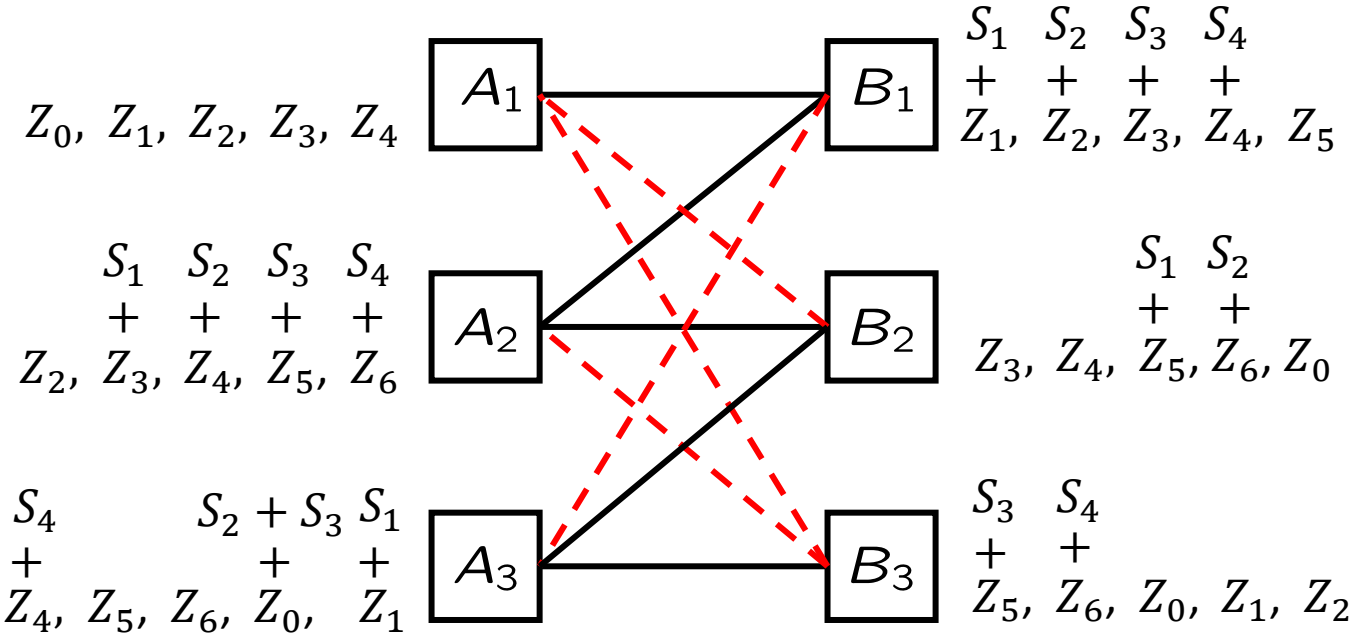}
\caption{\small A CDS instance contains an internal qualified edge $\{B_2, A_2\}$ located in an unqualified path $(B_2, A_1, B_3, A_2)$ within a qualified component $G_f$. The unqualified path has a distance $d = 3$, and the size of the connected edge cover is $\rho = 5$. The achievable noise rate is calculated as $(\rho - 1)(d - 1) / (\rho d - 1) = 4/7$. In this instance, the secret consists of $L = 4$ symbols, denoted as $S_1, S_2, S_3, S_4$, while the noise includes $L_Z = 7$ independent uniform symbols, represented as $Z_0, Z_1, \dots, Z_6$. Each message contains $N = 5$ symbols. Thus, the achieved rate is $R = L / L_Z = 4/7$.
}
\label{fig3} 
\end{center}
\end{figure}

\begin{example}[Upper bound for $\rho < +\infty$ ]
In Fig. \ref{fig3}, to cover all the nodes $A_2, B_3, A_1, B_2$ in residing unqualified path $(A_2, B_3, A_1, B_2)$, the connected edge cover should contain all the qualified edges in the graph $G_f$. The size of the connected edge cover is $\rho=5$ (see Definition \ref{def:cec}). The residing unqualified path distance is $d=3$ (see Definition \ref{def:rupd}). Then the linear noise rate upper bound is $R^{(\mbox{\scriptsize linear})}_Z \leq ((5-1)(3-1))/(3 \times 5-1)=4/7$. The achievable scheme for linear noise rate $4/7$ is shown in Fig. \ref{fig3}.
\hfill $\lozenge$
\end{example}

For the second example, there is no connected edge cover, i.e., $\rho=+\infty$.

\begin{example} [Upper bound for $\rho = +\infty$ ]
In Fig. \ref{fig2}, there is no connected edge cover, i.e., $\rho=+\infty$. The residing unqualified path distance is $d=3$ (see Definition \ref{def:rupd}). Then the linear noise rate upper bound is $R^{(\mbox{\scriptsize linear})}_Z \leq (d-1)/d=2/3$. 
In Fig. \ref{fig2}, we provide an achievable scheme achieving a linear noise rate of $2/3$, without restricting all nodes to have message size $N$.
\hfill $\lozenge$
\end{example}

\begin{remark}
In Theorem~2, we primarily aim to establish an upper bound. As this bound is not necessarily achievable in general, the achievable scheme presented here is included solely for illustration, to provide intuition for the upper bound.
\end{remark}

The above theorem demonstrates how the residing unqualified path distance governs the linear noise rate in a specific setting. Building on this intuition, we now introduce an additional requirement that the linear scheme must achieve the highest communication rate $N=L$. Under this stronger condition, a tighter linear noise rate upper bound can be derived, as stated in the following theorem.

\begin{theorem}\label{thm:linearnoiserate2}
\tit{For any CDS instance, if a linear coding scheme achieves the highest \comm rate, i.e., $N=L$, then the following linear noise rate upper bound holds:}
\begin{eqnarray}
R^{(\mbox{\scriptsize linear})}_Z \leq \frac{Q-1}{Q}
\end{eqnarray}
\end{theorem}

\begin{figure}[h]
\begin{center}
\includegraphics[width= 3 in]{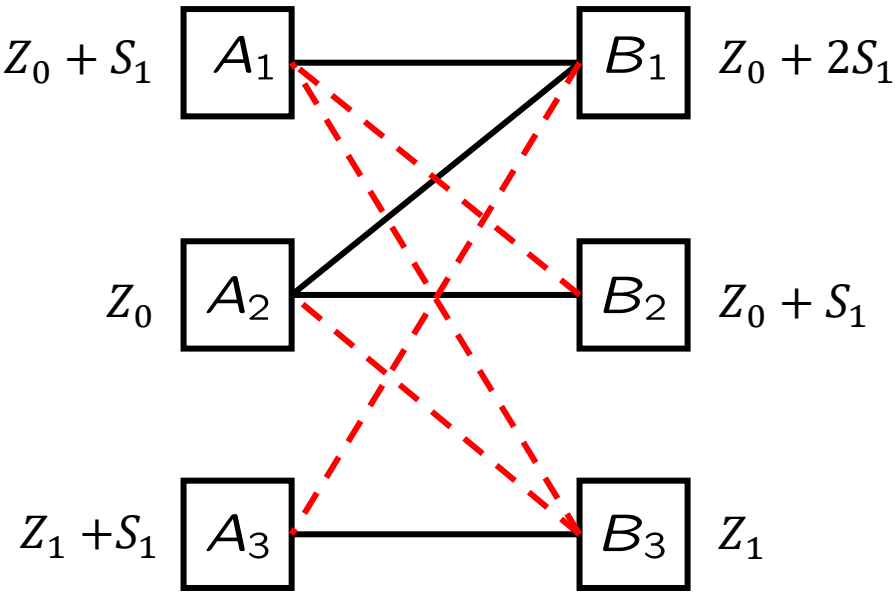}
\caption{\small A CDS instance contains an internal qualified edge $\{B_2, A_2\}$ located in an unqualified path $(B_2, A_1, B_3, A_2)$. The CDS scheme achieves the highest possible rate, i.e., $N = L$, with an achievable noise rate of $1/2$. In this scheme, the secret consists of $L = 1$ symbol, denoted as $S_1$, while the noise includes $L_Z = 2$ independent uniform bits, represented as $Z_0$ and $Z_1$. Each message contains $N = L = 1$ symbol. Thus, the linear noise rate achieved is $R^{(\mbox{\scriptsize linear})}_Z = L / L_Z = 1/2$.
}
\label{fig4}
\end{center}
\end{figure}

\begin{remark} 
For any CDS instance, if a linear coding scheme achieves the highest rate, i.e., $N = L$, then there are no internal qualified edges within a qualified component\footnote{For detailed proof, please refer to theorem 1 in \cite{Li_Sun_CDS}}, implying $\rho = +\infty$. According to Theorem \ref{thm:linearnoiserate1}, when $\rho = +\infty$, the linear noise rate is bounded above by $R^{(\mbox{\scriptsize linear})}_Z \leq (d-1)/d$.
Theorem \ref{thm:linearnoiserate2} introduces an additional constraint, $N = L$, which provides a tighter bound on the linear noise rate, specifically $R^{(\mbox{\scriptsize linear})}_Z \leq (Q-1)/Q$, where $Q \leq d$ and satisfies $(Q-1)/Q \leq (d-1)/d$ (Note that the definition of $Q$ is given in Definition~\ref{dec:crup}). 
\end{remark}

The proof of Theorem \ref{thm:linearnoiserate2} presented in section \ref{pfthm3}. Similar to Theorem \ref{thm:1noiserate},  we present one example with $Q=2$ to provide an intuitive understanding.

\begin{example} [Example for $N=L$]
In Fig. \ref{fig4}, consider the internal qualified edge $\{A_2, B_2\}$ and the residing unqualified path $(A_2, B_3, A_1,$ $B_2)$. Nodes $A_1, A_2, B_2$ belong to the same qualified component, while $B_3$ belongs to another qualified component. There are 2 qualified components connected through nodes located in the residing unqualified path, i.e., $Q=2$ (see definition \ref{dec:crup}).
Under the constraint $N = L$, nodes within the same qualified component must use the same noise. Based on this condition, the linear noise rate upper bound is $R^{(\mbox{\scriptsize linear})}_Z \leq (Q-1)/Q= 1/2$. The achievable scheme for the highest rate, $N = L$, with a linear noise rate of $1/2$, is illustrated in Fig. \ref{fig4}.
\hfill $\lozenge$
\end{example}

In the next theorem, we show that for a special class of CDS instances, the linear noise rate upper bound given in Theorem \ref{thm:linearnoiserate1} is in fact achievable. This result demonstrates that the bound is tight for these structured instances.

\begin{theorem} \label{thm:ach}
\tit{Consider any $(kd+1)+(kd+1)$ CDS instance%\footnote{A $k+k$ CDS instance refers to a bipartite graph with $k$ nodes in each part.}
, $k \in [K]$, where
\begin{enumerate}
    \item the qualified components consist of $(kd+1)$ cyclic qualified edges\footnote{Cyclic qualified edges are qualified edges that connect nodes in a cyclic manner between the two sets of a bipartite graph.}, and
    \item the unqualified edges form a path of unqualified path distance $d$.
\end{enumerate}
Then the linear noise capacity is
\[
R^{(\mbox{\scriptsize linear})}_Z = \frac{d-1}{d}.
\]
}
\end{theorem}

\begin{figure}[h]
\begin{center}
\includegraphics[width=.84\textwidth]{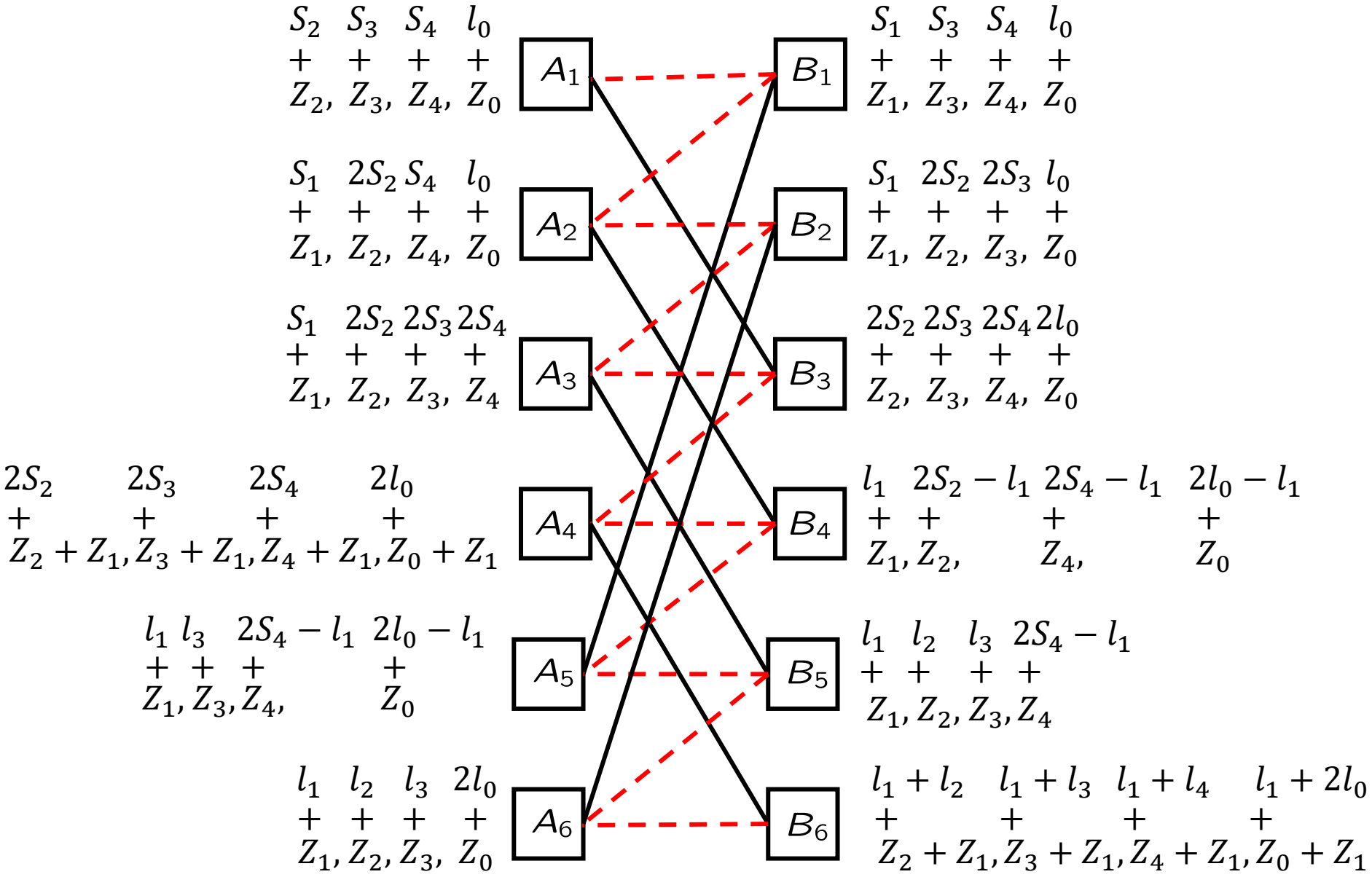}
\caption{\small 
A linear noise capacity achieving scheme for $4/5$. 
}
\label{fig5}
\end{center}
\end{figure}

The proof of Theorem \ref{thm:ach} is given in Section \ref{pfthm4}. For an intuitive understanding, we illustrate the scheme through an example.

\begin{example} [Example for the Achievability of Theorem \ref{thm:ach}]
In Fig. \ref{fig5}, consider a $6+6$ CDS instance with 6 nodes on each side. The 6 qualified components consist of cyclic qualified edges: $\{A_1,B_3\}, \{A_2,B_4\},$ $\{A_3,B_5\},$ $\{A_4,B_6\}, \{A_5,B_1\},$ $\{A_6,B_2\}$. The unqualified edges form a cyclic unqualified path $(\{A_1,B_1\},$ $\{B_1,A_2\},~ \{A_2,B_2\},~\{B_2,A_3\},\{A_3,B_3\},\{B_3,A_4\},~\{A_4,B_4\},~ \{B_4,A_5\}, \{A_5,B_5\}, \{B_5,A_6\}, \{A_6,B_6\})$. 
The residing unqualified path distance is $d=5$. This configuration satisfies the condition of Theorem \ref{thm:ach}, and the linear noise capacity is $R^{(\mbox{\scriptsize linear})}_Z = (d-1)/d = 4/5$. Next, we will illustrate the code assignment process.

First, we assign the noise variables. Consider the independent and identically distributed (i.i.d.) uniform noise variables \(Z_1, Z_2, Z_3, Z_4, Z_0\). Each node, \(A_1, B_1, A_2, B_2, A_3\), is assigned \(d-1=4\) noise symbols from $Z_1, Z_2, Z_3,$ $ Z_4,$ $ Z_0$ in a circular manner, as shown in Fig. \(\ref{fig5}\). \(\{A_1, B_3\}\) is a qualified edge, meaning that the secret can be recovered using the nodes \(A_1\) and \(B_3\). To ensure this, \(B_3\) must be assigned the same noise symbols as \(A_1\). Therefore, node \(B_3\) is assigned \(Z_0, Z_2, Z_3, Z_4\).
Node \(A_4\), in turn, is assigned the noise symbols \(Z_0+Z_1, Z_2+Z_1, Z_3+Z_1, Z_4+Z_1\). For the remaining nodes \(B_4, A_5, B_5, A_6, B_6\), the correctness constraint ensures that each node is assigned the same noise symbols as its corresponding counterpart in a qualified edge. 
With these assignments, the noise distribution for all nodes is complete, as illustrated in Fig. \(\ref{fig5}\).

Second, we assign the secret. Consider the nodes \(A_1, B_1, A_2, B_2, A_3, B_3\). We assign the symbols \(S_1, S_2, S_3, S_4, l_0\) (where \(l_i\) represents one linear combination of the secret symbols \(S_1, S_2, S_3, S_4\), as denoted in Fig.~\(\ref{fig5}\)) to the noise variables \(Z_1, Z_2, Z_3, Z_4, Z_0\), respectively. 

Next, we describe the assignment of coefficients. For each noise symbol \(Z_i, i \in \{0,1,\cdots,4\}\), we consider the nodes that contain \(Z_i\) and examine the unqualified edges associated with \(Z_i\). For every unqualified path involving \(Z_i\), the same coefficient (i.e., the same message) is assigned. For different unqualified paths, distinct coefficients (i.e., different messages) are assigned. 

For example, consider the noise \(Z_4\). Nodes \(A_1, B_1, A_2\) lie along the same unqualified path, so the noise \(Z_4\) in these nodes is assigned the same message, \(S_4 + Z_4\). Nodes \(A_3, B_3\), which belong to a different unqualified path, are assigned a different message, \(2S_4 + Z_4\). Note that node \(B_2\) is not included in any unqualified path for \(Z_4\) because \(Z_4\) is not present in \(B_2\).

Now, consider the secrets in \(A_4\). To satisfy the security constraint for the unqualified edge \(\{A_4, B_3\}\), the same secrets must be assigned to \(A_4\) as in \(B_3\). For instance, we assign the secrets \(2l_0, 2S_2, 2S_3, 2S_4\) to the noises \(Z_0 + Z_1, Z_2 + Z_1, Z_3 + Z_1, Z_4 + Z_1\), ensuring that no information can be revealed from \(\{A_4, B_3\}\). For the secrets in \(B_4\), we need to address both the correctness constraint for \(\{A_2, B_4\}\) and the security constraint for \(\{A_4, B_4\}\). For the correctness constraint, nodes \(B_4\) and \(A_2\) share the same noise, so the secrets in \(B_4 - A_2\) must be carefully designed to be linearly independent. This ensures that the secret matrix corresponding to \(B_4 - A_2\) is full rank, allowing the secrets to be recovered from \(A_2\) and \(B_4\). For the security constraint, we assign the secret \(l_1\) to the noise \(Z_1\), resulting in the symbol \(l_1 + Z_1\). If the noise at any node in an unqualified edge is a linear combination of the noise at another node, the secret associated with this noise must also be a linear combination of the secret at the other node. For example, consider the symbols containing \(Z_1\) and \(Z_2\) in nodes \(A_4\) and \(B_4\). The symbol \(2S_2 - l_1 + Z_2\) in node \(B_4\) must be a linear combination of the symbols containing \(Z_1\) and \(Z_2\). Hence, we assign the secret \(2S_2 - l_1\) to the noise \(Z_2\) in node \(B_4\). Following this logic, we similarly assign secrets to the noises \(Z_4\) and \(Z_0\) as shown in Fig.~\ref{fig5}.

Next, consider the nodes \(A_5, B_5, A_6\). We examine each noise symbol \(Z_i, i \in \{0,1,\cdots,4\}\) sequentially and the unqualified edges associated with each noise symbol. For each noise symbol \(Z_i\), if it appears in the same unqualified path as in node \(B_4\), we assign the same message as in \(B_4\). For example, consider the noise \(Z_0\). Since \(A_5\) lies along the same unqualified path as \(B_4\), the message associated with \(Z_0\) in \(A_5\) is the same as in \(B_4\), i.e., \(2l_0 - l_1 + Z_0\). However, \(A_6\) does not lie on the same unqualified path as \(B_4\), so the message associated with \(Z_0\) in \(A_6\) is assigned as \(2l_0\) (since \(B_5\) does not contain \(Z_0\)).

Finally, consider \(B_6\). The secrets of \(Z_0 + Z_1\) are linear combinations of the secrets for \(Z_0\) and \(Z_1\) in \(A_6\). The same logic applies to \(Z_2 + Z_1\) and \(Z_3 + Z_1\). Additionally, we assign \(l_1 + l_4\) to \(Z_4 + Z_1\). With this, the code assignment is complete.

The scheme is secure because the assignment satisfies the security constraint (\ref{eq:message}) for all unqualified edges. In fact, the assignment is directly guided by (\ref{eq:message}). For example, consider the unqualified edge \(\{A_4, B_4\}\). The message \(2S_2 + Z_2 + Z_1\) is a linear combination of \(l_1 + Z_1\) and \(2S_2 - l_1 + Z_2\), ensuring that no information about the secret is revealed. A similar logic applies to other symbols. For correctness, we observe that any two nodes in a qualified edge share the same noise. For qualified edges before node \(B_3\), the secret symbols are assigned distinct coefficients. For example, from \(\{A_1, B_3\}\), we can recover \((l_0, S_2, S_3, S_4)\). For qualified edges connected to nodes after \(B_3\), the secrets are carefully designed to maintain security.
\hfill $\lozenge$
\end{example}

\section{Proof of Theorem \ref{thm:1noiserate}}\label{pfthm1}

\subsection{The ``Only if" Part}
\label{sec:thm11}
Consider any CDS instance where each node is connected to at least one unqualified edge, as described by the characteristic graph $G_f(V,E)$. We show that if the feasibility condition in Theorem~\ref{thm:1noiserate} is violated, a noise rate of $R_Z = 1$ cannot be achieved. Without loss of generality, assume there exist at least one internal qualified edge $\{v_1,v_P\}$ and residing unqualified path $(\{v_1, v_2\}, \{v_2, v_3\}, $ $\cdots,$ $ \{v_{P-1}, v_P\})$. To establish this, we use a proof by contradiction. Assume that $R_Z = \lim_{L \to \infty} \frac{L}{L_Z} = 1$ is achievable, implying $L_Z = L + o(L)$. Consequently, for any message connected to a qualified edge, the entropy of the noise used in such a message must also satisfy $H(Z) = L + o(L)$. This conclusion is formalized in Lemma \ref{lemma:size}.

\begin{lemma}[Message and Noise Size] \label{lemma:size}
\tit{When $R_Z = 1$, for any message $v \in V$ such that there exists $u \in V$ such that $\{v,u\}$ is a qualified edge, we have}
\begin{eqnarray}
H(v) = H(v | S) =H(Z)= L+o(L). \label{eq:size}
\end{eqnarray}
\end{lemma}

\begin{IEEEproof}
For any node $w$, there exists a node $w'$ such that $\{w, w'\}$ is unqualified. From the security constraint (\ref{sec}), we have
\begin{eqnarray}
I(w, w'; S) = 0 &\Rightarrow& I(w; S) = 0 \\
\mbox{($w$ can by any node)} &\Rightarrow& I(v; S) = I(u;S) = 0. \label{eq:usind}
\end{eqnarray}
Consider now the qualified edge $\{v,u\}$. From the correctness constraint (\ref{dec}), we have
\begin{eqnarray}
L &\overset{(\ref{sz_ind})}{=}& H(S) \overset{(\ref{dec})}{=} I(v,u; S) \overset{(\ref{eq:usind})}{=} I(v; S|u)\leq H(v) \label{leq:ln}\\
&\overset{(\ref{eq:usind})}{=}& H(v|S)=H(v|S,Z)+ I(v;Z|S)\\
&\overset{(\ref{det})}{\leq}& H(Z|S)\leq H(Z)= L_Z=L+o(L).
\end{eqnarray}
The proof is thus complete.
\end{IEEEproof}

Next, we show that the joint entropy of all nodes $V$ is $L + o(L)$, which is approximately equal to the entropy of the noise observed at each individual node. In other words, the noise across all nodes must be fully aligned.

\begin{lemma}[Noise Alignment for all nodes $V$] \label{lemma:noise1}
\tit{When $R_Z = 1$, for all nodes $V$, we have}
\begin{eqnarray}
H(V | S) = L+o(L). \label{eq:component}
\end{eqnarray}
\end{lemma}

\begin{IEEEproof} 
On the one hand, we have
\begin{eqnarray}
H(V | S) &=& H(V|Z, S) +I(V;Z|S) \overset{(\ref{dec})}{\leq} H(Z|S) \leq H(Z)=L_Z=L+o(L).
\end{eqnarray}
On the other hand, we have
\begin{eqnarray}
H(V | S) \geq H(v|S) &\overset{(\ref{eq:size})}{=}& L +o(L).
\end{eqnarray}
The proof is now complete.
\end{IEEEproof}

We now proceed to the message alignment phenomenon. We show that any two vertices $v,u$ that form an unqualified edge must produce identical message. In other words, the joint entropy of $v, u$ is $L+o(L)$, the same as that of any individual $v$ or $u$. 

\begin{lemma}[Message Alignment for Unqualified Edge] \label{lemma:message1}
\tit{When $R_Z = 1$, for any unqualified edge $\{v,u\}$, we have}
\begin{eqnarray}
H(v,u) = L+o(L). \label{eq:unedge}
\end{eqnarray}
\end{lemma}

\begin{IEEEproof}
Note that both end nodes of the unqualified edge $\{v,u\}$ belong to the node set of $V$. Combining the security constraint (\ref{sec}) and (\ref{eq:component}), we have
\begin{eqnarray}
 L+o(L)\overset{(\ref{eq:size})}{=}H(v)\leq H(v,u) \overset{(\ref{sec})}{=} H(v,u | S)\leq H(V|S) \overset{(\ref{eq:component})}{=} L+o(L).
\end{eqnarray}
\end{IEEEproof}

In the following lemma, we generalize the message alignment phenomenon from unqualified edges to unqualified paths.

\begin{lemma}[Message Alignment for Unqualified Path] \label{lemma:message2}
\tit{When $R_Z = 1$, for any unqualified path, $(\{v_1, v_2\}$, $\{v_2, v_3\}, $ $\cdots,$ $ \{v_{P-1}, v_P\})$, we have}
\begin{eqnarray}
H(v_1, v_P) \leq L+o(L). \label{eq:unpath}
\end{eqnarray}
\end{lemma}

\begin{IEEEproof} 
Equipped with what has been established, the proof follows from a simple recursive application of the sub-modularity property of entropy functions.
\begin{eqnarray}
(P-1) L+o(L)&\overset{(\ref{eq:unedge})}{=}&H(v_1, v_2) + H(v_2, v_3) + \cdots H(v_{P-1}, v_P)\\
&\geq& H(v_1, v_2, \cdots, v_P) + H(v_2) +  \cdots + H(v_{P-1}) \notag\\
&\overset{(\ref{eq:size})}{\geq}& H(v_1, v_P) + (P-2)L +o(L) \notag\\
\Rightarrow ~~~~~~~H(v_1, v_P) &\leq& L +o(L).
\end{eqnarray}
\end{IEEEproof}

After establishing the above lemmas, we are now ready to identify the contradiction. Specifically, since the noise capacity condition of $1$ is violated, there must exist an internal qualified edge (denoted as $\{v_1, v_P\}$) within a residing unqualified path, which can be expressed as $(\{v_1, v_2\}, \{v_2, v_3\}, \dots, \{v_{P-1}, v_P\})$. According to the correctness constraint (\ref{dec}) of the qualified edge $\{v_1, v_P\}$, we have
\begin{eqnarray}
L +o(L) \overset{(\ref{eq:unpath})}{\geq} H(v_1, v_P) \overset{(\ref{dec})}{=} H(v_1, v_P, S) \geq H(S) + H(v_1| S) \overset{(\ref{sz_ind})(\ref{eq:unedge})}{=}  L + L +o(L). \label{eq:nn}
\end{eqnarray}
So normalizing (\ref{eq:nn}) by $L$ and letting $L$ approach infinity, we have $1 \geq 2$, and the contradiction is arrived.
The proof of the only if part is thus complete.

\subsection{The ``If"  Part}
\label{sec:if}
We demonstrate that if the condition for $R_Z = 1$ in Theorem \ref{thm:1noiserate} is satisfied, the CDS noise capacity equals 1. To establish this, we first prove that $R_Z \leq 1$, and then show that $R_Z = 1$ is achievable.

The proof of $R_Z \leq 1$ is as follows. Consider any CDS instance that contains at least one qualified edge $\{v,u\}$; otherwise all edges are unqualified, the problem is meaningless as the secret is never disclosed. Further, each node is connected to an unqualified edge $\{u,w\}$. From the security constraint (\ref{sec}), we have
\begin{eqnarray}
I(u,w ; S) = 0 &\Rightarrow& I(u;S) = 0. \\
\mbox{($u$ can by any vertex)} &\Rightarrow& I(v; S) = I(u;S) = 0.\label{eq:i11}
\end{eqnarray}
From the correctness constraint (\ref{dec}), we have
\begin{eqnarray}
&& L \overset{(\ref{sz_ind})}{=} H(S) \overset{(\ref{dec})}{=} I(S; v,u) \overset{(\ref{eq:i11})}{=} I(S; v | u) \leq H(v)\label{leq:ln2}\\
&\overset{(\ref{eq:i11})}{=}&H(v|S)=H(v|S,Z)+I(v;Z|S)\overset{(\ref{det})}{\leq} H(Z|S)\leq H(Z) \overset{(\ref{sz_ind})}{=} L_Z \\
&\Rightarrow& R_Z = L/L_Z \leq 1.
\end{eqnarray}}

We now present the coding scheme that achieves noise rate $1$. The scheme is a generalization of that presented in Example \ref{ex1}.

Consider any CDS instance where each node is connected to at least one qualified edge and one unqualified edge, described by the characteristic graph $G_f(V,E)$. Suppose $G_f(V,E)$ has  $U$ unqualified components.
Choose $p$ as a prime number that is no fewer than $U$. The secret $S$ contains $L = 1$ symbol from the finite field $\mathbb{F}_p$ and the noise $Z$ contains $L_Z = 1$ symbols from $\mathbb{F}_p$.

The messages are assigned as follows:
\begin{eqnarray}
\mbox{We set  any message $v$ in the $i^{th}, i \in \{1,2,\cdots,U\}$ unqualified component as} ~ Z + i S. \label{eq:scheme}
\end{eqnarray}

To complete the proof of the achievable scheme, we demonstrate that the scheme is both correct and secure. We begin with the correctness constraint (\ref{dec}). Since the noise capacity condition of $R_Z=1$ in Theorem \ref{thm:1noiserate} is satisfied, there are no internal qualified edges; in other words, any qualified edge must connect nodes belonging to different unqualified components. Consider any qualified edge $\{v, u\}$, where $v$ belongs to the $i^{\text{th}}$ unqualified component and $u$ belongs to the $j^{\text{th}}$ unqualified component. Note that $j \neq i$. From (\ref{eq:scheme}), we have
\begin{eqnarray}
&& v = Z + i S, u = Z + j S \\
&\Rightarrow& H(S | v,u)  = H(S | Z + i S, Z + j S) \overset{j \neq i}{=} H(S | S, Z) = 0
\end{eqnarray}
so that the scheme is always correct.

Next consider the security constraint (\ref{sec}). Any unqualified edge must belong to the same unqualified component. Consider any unqualified edge $\{v,u\}$. We have
\begin{eqnarray}
&& v = u = Z + i S \label{eq:zz1} \\
&\Rightarrow& H(S | v,u)  = H(S | Z + i S) = H(S, Z + i S) - H(Z + i S) = 1 = H(S)
\end{eqnarray}
so that security is guaranteed.

\section{Proof of Theorem \ref{thm:linearnoiserate1}}\label{pfthm2}
The proof of Theorem \ref{thm:linearnoiserate1} follows similarly from that of the CDS instance in Fig.~\ref{fig3} considered in the previous section. We first simplify a notation that will be frequently used. For nodes $v_1, \cdots, v_i$, denote the dimension of the overlap of their noise spaces as $\alpha_{v_1\cdots v_i}$, i.e.,
\begin{eqnarray}
\alpha_{v_1\cdots v_i} \triangleq \dim( \mbox{rowspan}({\bf H}_{v_1}) \cap \cdots \cap \mbox{rowspan}({\bf H}_{v_i}) ). \label{eq:a12}
\end{eqnarray}

For a single node $v$, denote the dimension of the noise space as $r_v$, i.e., 
\begin{eqnarray}
r_{v} \triangleq \dim( \mbox{rowspan}({\bf H}_{v}))=N. \label{eq:rv}
\end{eqnarray}

Next, we consider two cases, $\rho < +\infty$ and $\rho < +\infty$. For the first case, we proof that $R^{(\mbox{\scriptsize linear})}_Z \leq ((\rho-1)(d-1))/(\rho d-1)$. For the second case, we proof that $R^{(\mbox{\scriptsize linear})}_Z \leq (d-1)/d$. Consider the first case, we first proof that $N\geq (\rho L)/(\rho-1)$.
This result has appeared as theorem 1 in \cite{Li_Sun_linearCDS} and a proof is presented here for completeness.

Consider any CDS instance $G_f(V,E)$, where $\rho < +\infty$ and focus on an internal qualified edge $e$ in a residing unqualified path $P$ such that $\rho(e, P) = \rho$. Then the connected edge cover $M$ for nodes $V_P$ in $P$ contains $\rho$ edges and $\rho+1$ nodes, denoted as $V_M = \{v_1, v_2, \cdots, v_{\rho+1}\} \subset V$. Note that such $e, P, M$ are guaranteed to exist as $\rho < +\infty$ and according to the definition of $\rho$, the connected edge cover $M$ attains the minimal cardinality so that $M$ is a spanning tree of the nodes $V_M$.

Start with the internal qualified edge $e$ in $M$, say $e = \{v_{i_1}, v_{i_2}\} \subset M, i_1, i_2 \in \{1,2, \cdots, \rho+1\}$. 
As $M$ is connected, there must exist a node $v_{i_3} \in V_M, i_3 \notin \{i_1, i_2\}$ and a node $u_1 \in \{v_{i_1}, v_{i_2}\}$ such that $\{u_1, v_{i_3}\}$ is a qualified edge. Then from sub-modularity, we have
\begin{eqnarray}
\alpha_{v_{i_1}v_{i_2}v_{i_3}} 
\geq \alpha_{v_{i_1}v_{i_2}} + \alpha_{u_1v_{i_3}} - N. \label{eq:a123}
\end{eqnarray}

Then we proceed similarly to find $v_{i_4} \in V_M, i_4 \notin \{i_1, i_2, i_3\}$ such that $\{u_2, v_{i_4}\}$ is a qualified edge, where $u_2 \in \{v_{i_1}, v_{i_2}, v_{i_3}\}$. Again from sub-modularity, we have
\begin{eqnarray}
\alpha_{v_{i_1}v_{i_2}v_{i_3}v_{i_4}} &\geq& \alpha_{v_{i_1}v_{i_2}v_{i_3}} + \alpha_{u_2v_{i_4}} - N \\
&\overset{(\ref{eq:a123})}{\geq}& \alpha_{v_{i_1}v_{i_2}} + \alpha_{u_1v_{i_3}} + \alpha_{u_2v_{i_4}} - 2N.
\end{eqnarray}
Continue this procedure, i.e., we include one node $v_{i_j} \in V_M, i_j \notin \{i_1, \cdots, i_{j-1}\}, j \in \{5,\cdots, \rho+1\}$ at one time such that $\{u_{j-2}, v_{i_{j}}\} \in M$ and $u_{j-2} \in \{v_{i_1}, \cdots, v_{i_{j-1}}\}$. Then we have
\begin{eqnarray}
\alpha_{v_{i_1}v_{i_2}\cdots v_{i_{\rho+1}}} &\geq& \alpha_{v_{i_1}\cdots v_{i_\rho}} + \alpha_{u_{\rho-1} v_{i_{\rho+1}}} - N \\
&\geq& \cdots \\
&\geq& \alpha_{v_{i_1}v_{i_2}} + \alpha_{u_1v_{i_3}} + \alpha_{u_2v_{i_4}} + \cdots + \alpha_{u_{\rho-1}v_{i_{\rho+1}}} - (\rho-1)N.  
\label{eq:v1v2v3vp} 
\end{eqnarray}
Note that $i_1, \cdots, i_{\rho+1}$ are distinct so that $V_M = \{v_1, \cdots, v_{\rho+1}\} = \{v_{i_1}, \cdots, v_{i_{\rho+1}}\}$.

As the $\rho+1$ noise spaces have an overlap of dimension $\alpha_{v_{i_1}v_{i_2}\cdots v_{i_{\rho+1}}}$, there exist $\rho+1$ projection matrices ${\bf P}^{\cap}_{v_{i_1}}, \cdots, {\bf P}^{\cap}_{v_{i_{\rho+1}}}$ of rank $\alpha_{v_{i_1}v_{i_2}\cdots v_{i_{\rho+1}}}$ each such that
\begin{eqnarray}
&&{\bf P}^{\cap}_{v_{i_1}}{\bf H}_{v_{i_1}} = {\bf P}^{\cap}_{v_{i_2}}{\bf H}_{v_{i_2}} = \cdots = {\bf P}^{\cap}_{v_{i_{\rho+1}}}{\bf H}_{v_{i_{\rho+1}}},\notag\\
&&\mbox{rank}({\bf P}^{\cap}_{v_{i_1}}) =\cdots = \mbox{rank}({\bf P}^{\cap}_{v_{i_{\rho+1}}})= \mbox{dim}(\mbox{rowspan}({\bf H}_{v_{i_1}}) \cap\cdots \cap \mbox{rowspan}({\bf H}_{v_{i_{\rho+1}}}))\label{eq:s1}
\end{eqnarray} 

Next, switch focus to the unqualified path $P$. Consider the nodes $V_P \subset V_M$ and denote $V_P =\{v_{i_1}, v_{j_1}, v_{j_2}, \cdots, $ $ v_{j_{d-1}}, v_{i_{2}}\} \subset \{v_{i_1}, v_{i_2}, \cdots, v_{i_{\rho+1}}\} = V_M$ such that $\{v_{i_1}, v_{j_1}\}$, $\{v_{j_1}, v_{j_2}\}$, $\cdots$, $\{v_{j_{d-1}}, v_{i_2}\}$ are unqualified edges. By (\ref{eq:message}), i.e., the message alignment constraint from Lemma \ref{lemma:align}, and (\ref{eq:s1}), we have
\begin{eqnarray}
&& {\bf P}^{\cap}_{v_{i_1}}{\bf F}_{v_{i_1}} = {\bf P}^{\cap}_{v_{j_1}}{\bf F}_{v_{j_1}} = \cdots = {\bf P}^{\cap}_{v_{j_{d-1}}}{\bf F}_{v_{j_{d-1}}} = {\bf P}^{\cap}_{v_{i_2}}{\bf F}_{v_{i_2}} \notag \\
&\Rightarrow&{\bf P}^{\cap}_{v_{i_1}}{\bf F}_{v_{i_1}} = {\bf P}^{\cap}_{v_{i_2}}{\bf F}_{v_{i_2}}.
\label{eq:v1v2s}
\end{eqnarray}

Finally, consider the internal qualified edge $e = \{v_{i_1}, v_{i_2}\}$ and identify the noise overlap through matrices ${\bf P}_{v_{i_1}}, {\bf P}_{v_{i_2}}$ that have rank $\alpha_{v_{i_1}v_{i_2}}$, i.e., ${\bf P}_{v_{i_1}} {\bf H}_{v_{i_1}} = {\bf P}_{v_{i_2}} {\bf H}_{v_{i_2}}$. Noting that $\mbox{rowspan}({\bf P}^\cap_{v_{i_1}})$ is a subspace of $\mbox{rowspan}({\bf P}_{v_{i_1}})$, we set
\begin{eqnarray}
{\bf P}^\cap_{v_{i_1}} = {\bf P}_{v_{i_1}}(1: \alpha_{v_{i_1}v_{i_2}\cdots v_{i_{\rho+1}}}, :), ~~{\bf P}^\cap_{v_{i_2}} = {\bf P}_{v_{i_2}}(1: \alpha_{v_{i_1}v_{i_2}\cdots v_{i_{\rho+1}}}, :) \label{eq:s2s}
\end{eqnarray}
without loss of generality, i.e., the first $\alpha_{v_{i_1}v_{i_2}\cdots v_{i_{\rho+1}}}$ rows of ${\bf P}_{v_{i_1}}$ are ${\bf P}^\cap_{v_{i_1}}$.
Then from the correctness constraint (\ref{eq:noise}) for qualified edge $e = \{v_{i_1}, v_{i_2}\}$, we have
\begin{eqnarray}
L &\overset{(\ref{eq:dec})}{\leq}&\mbox{rank}\left({\bf P}_{v_{i_1}} {\bf F}_{v_{i_1}} - {\bf P}_{v_{i_2}} {\bf F}_{v_{i_2}}\right) \\
&\overset{(\ref{eq:v1v2s}) (\ref{eq:s2s})}{=}& \mbox{rank} \left( {\bf P}_{v_{i_1}} (\alpha_{v_{i_1}v_{i_2}\cdots v_{i_{\rho+1}}} +1:\alpha_{v_{1}v_{2}}, :) {\bf F}_{v_{i_1}} - {\bf P}_{v_{i_2}}(\alpha_{v_{i_1}v_{i_2}\cdots v_{i_{\rho+1}}} +1:\alpha_{v_{1}v_{2}}, :) {\bf F}_{v_{i_2}} \right) \notag\\
&&\\
&\leq& \alpha_{v_{i_1}v_{i_2}} - \alpha_{v_{i_1}v_{i_2}\cdots v_{i_{\rho+1}}} \\
&\overset{(\ref{eq:v1v2v3vp})}{\leq}&\alpha_{v_{i_1}v_{i_2}} - \left( \alpha_{v_{i_1}v_{i_2}} + \alpha_{u_1v_{i_3}} + \alpha_{u_2v_{i_4}} + \cdots + \alpha_{u_{\rho-1}v_{i_{\rho+1}}} - (\rho-1)N \right)\\
&=& (\rho-1)N - \left(\alpha_{u_1v_{i_3}} + \alpha_{u_2v_{i_4}} + \cdots + \alpha_{u_{\rho-1}v_{i_{\rho+1}}} \right) \\
&\overset{(\ref{eq:noise})}{\leq}& (\rho-1)N - (\rho-1)L \\
& \Rightarrow & N\geq \frac{\rho}{\rho-1} L \label{geq:npl}
\end{eqnarray}

Consider the second case, i.e., $\rho=+\infty$, we proof that $N\geq L$. Consider (\ref{leq:ln2}), we have
\begin{eqnarray}
    L &\overset{(\ref{sz_ind})}{=}& H(S) \overset{(\ref{dec})}{=} I(v,u; S) \overset{(\ref{eq:i11})}{=} I(v; S|u)\leq H(v)\overset{(\ref{messagezise})}{=}N\\
    \Rightarrow&&N\geq L\label{geq:nl}
\end{eqnarray}

Next, consider the residing unqualified path $V_P =\{v_{i_1}, v_{j_1}, v_{j_2}, \cdots, v_{j_{d-1}}, v_{i_{2}}\}$. For the noise space overlap of any two nodes, say $v_{i_1}, v_{j_1}$, the total noise space is $L_Z$, then by sub-modularity, we have
\begin{eqnarray}
    \alpha_{v_{i_1}v_{j_1}}\geq r_{v_{i_1}}+r_{v_{j_1}}-L_Z.\label{geq:lzs}
\end{eqnarray}

Then we proceed similarly for the noise space overlap of any three nodes, say $v_{i_1}, v_{j_1}, v_{j_2}$. Again from sub-modularity, we have
\begin{eqnarray}
    \alpha_{v_{i_1}v_{j_1}v_{j_2}}&\geq& \alpha_{v_{i_1}v_{j_1}}+\alpha_{v_{j_1}v_{j_2}}-N\\
    &\geq& \alpha_{v_{i_1}v_{j_1}}+r_{v_{j_1}}+r_{v_{j_2}}-L_Z-N\\
    &\geq& r_{v_{i_1}}+2r_{v_{j_1}}+r_{v_{j_2}}-2L_Z-N
\end{eqnarray}

Consider the noise space overlap of any $d+1$ nodes, say $v_{i_1}, v_{j_1}, v_{j_2}, \cdots, v_{j_{d-1}}, v_{i_{2}}$, we have
\begin{eqnarray}
    \alpha_{v_{i_1}v_{j_1}v_{j_2}\cdots v_{j_{d-1}}v_{i_{2}}}&\geq&  \alpha_{v_{i_1}v_{i_{2}}}+\alpha_{v_{i_1}v_{j_1}}+\alpha_{v_{j_1}v_{j_2}}+\cdots+\alpha_{v_{j_{d-2}}v_{j_{d-1}}}-(d-1)N\label{eq:dp1}\\ 
    &\geq&  \alpha_{v_{i_1}v_{i_{2}}}+r_{v_{i_1}}+2r_{v_{j_1}}+2r_{v_{j_2}}+\cdots+2r_{v_{j_{d-2}}}+r_{v_{j_{d-1}}}\notag\\
    &&-(d-1)L_Z-(d-1)N\label{leq:lzn}
\end{eqnarray}

As the $d+1$ noise spaces have an overlap of dimension $\alpha_{v_{i_1}v_{j_1}v_{j_2}\cdots v_{j_{d-1}}v_{i_{2}}}$, there exist $d+1$ projection matrices ${\bf P}^{\cap}_{v_{i_1}},{\bf P}^{\cap}_{v_{j_1}}, \cdots, {\bf P}^{\cap}_{v_{i_{d-1}}},{\bf P}^{\cap}_{v_{i_2}}$ of rank $\alpha_{v_{i_1}v_{j_1}v_{j_2}\cdots v_{j_{d-1}}v_{i_{2}}}$ each such that
\begin{eqnarray}
&&{\bf P}^{\cap}_{v_{i_1}}{\bf H}_{v_{i_1}} = {\bf P}^{\cap}_{v_{j_1}}{\bf H}_{v_{j_1}} = \cdots = {\bf P}^{\cap}_{v_{j_{d-1}}}{\bf H}_{v_{j_{d-1}}}={\bf P}^{\cap}_{v_{i_1}}{\bf H}_{v_{i_2}},\notag\\
&&\mbox{rank}({\bf P}^{\cap}_{v_{i_1}}) =\cdots = \mbox{rank}({\bf P}^{\cap}_{v_{i_{2}}})= \mbox{dim}(\mbox{rowspan}({\bf H}_{v_{i_1}}) \cap\cdots \cap \mbox{rowspan}({\bf H}_{v_{i_{2}}}))\label{eq:s1s2}\\
\overset{(\ref{eq:message})}{\Rightarrow}&&{\bf P}^{\cap}_{v_{i_1}}{\bf F}_{v_{i_1}} = {\bf P}^{\cap}_{v_{j_1}}{\bf F}_{v_{j_1}} = \cdots = {\bf P}^{\cap}_{v_{j_{d-1}}}{\bf F}_{v_{j_{d-1}}}={\bf P}^{\cap}_{v_{i_2}}{\bf F}_{v_{i_2}}\\
\Rightarrow&&{\bf P}^{\cap}_{v_{i_1}}{\bf F}_{v_{i_1}} ={\bf P}^{\cap}_{v_{i_2}}{\bf F}_{v_{i_2}}\label{eq:st1}
\end{eqnarray} 

Finally, consider the internal qualified edge $e = \{v_{i_1}, v_{i_2}\}$ and identify the noise overlap through matrices ${\bf P}_{v_{i_1}}, {\bf P}_{v_{i_2}}$ that have rank $\alpha_{v_{i_1}v_{i_{2}}}$, i.e., ${\bf P}_{v_{i_1}} {\bf H}_{v_{i_1}} = {\bf P}_{v_{i_2}} {\bf H}_{v_{i_2}}$. Noting that $\mbox{rowspan}({\bf P}^\cap_{v_{i_1}})$ is a subspace of $\mbox{rowspan}({\bf P}_{v_{i_1}})$, we set
\begin{eqnarray}
{\bf P}^\cap_{v_{i_1}} = {\bf P}_{v_{i_1}}(1: \alpha_{v_{i_1}v_{j_1}v_{j_2}\cdots v_{j_{d-1}}v_{i_{2}}}, :), ~~{\bf P}^\cap_{v_{i_2}} = {\bf P}_{v_{i_2}}(1: \alpha_{v_{i_1}v_{j_1}v_{j_2}\cdots v_{j_{d-1}}v_{i_{2}}}, :) \label{eq:st2}
\end{eqnarray}
without loss of generality, i.e., the first $\alpha_{v_{i_1}v_{j_1}v_{j_2}\cdots v_{j_{d-1}}v_{i_{2}}}$ rows of ${\bf P}_{v_{i_1}}$ are ${\bf P}^\cap_{v_{i_1}}$.
Then from the correctness constraint (\ref{eq:dec}) for qualified edge $e = \{v_{i_1}, v_{i_2}\}$, we have
\begin{eqnarray}
L &\overset{(\ref{eq:dec})}{\leq}&\mbox{rank}\left({\bf P}_{v_{i_1}} {\bf F}_{v_{i_1}} - {\bf P}_{v_{i_2}} {\bf F}_{v_{i_2}}\right) \\
&\overset{(\ref{eq:st1}) (\ref{eq:st2})}{=}& \mbox{rank} \left( {\bf P}_{v_{i_1}} (\alpha_{v_{i_1}v_{j_1}v_{j_2}\cdots v_{j_{d-1}}v_{i_{2}}} +1:\alpha_{v_{i_1}v_{i_2}}, :) {\bf F}_{v_{i_1}}\right. \notag\\
&&-\left. {\bf P}_{v_{i_2}}(\alpha_{v_{i_1}v_{j_1}v_{j_2}\cdots v_{j_{d-1}}v_{i_{2}}} +1:\alpha_{v_{i_1}v_{i_2}}, :) {\bf F}_{v_{i_2}} \right) \\
&\leq& \alpha_{v_{i_1}v_{i_2}} - \alpha_{v_{i_1}v_{j_1}v_{j_2}\cdots v_{j_{d-1}}v_{i_{2}}} \\
&\overset{(\ref{leq:lzn})}{\leq}&\alpha_{v_{i_1}v_{i_2}} - \big(\alpha_{v_{i_1}v_{i_{2}}}+r_{v_{i_1}}+2r_{v_{j_1}}+2r_{v_{j_2}}+\cdots+2r_{v_{j_{d-2}}}+r_{v_{j_{d-1}}}\notag\\
    &&-(d-1)L_Z-(d-1)N\big)\\
&=&- \left( r_{v_{i_1}}+2r_{v_{j_1}}+2r_{v_{j_2}}+\cdots+2r_{v_{j_{d-2}}}+r_{v_{j_{d-1}}}-(d-1)L_Z-(d-1)N \right)\\
&\overset{(\ref{eq:rv})}{=}&-(2(d-1)N-(d-1)L_Z- (d-1)N)\\
&=&(d-1)L_Z- (d-1)N\label{leq:llzn}
\end{eqnarray}

For the first case, i.e., $\rho< +\infty$, from (\ref{geq:npl})(\ref{leq:llzn}), we have
\begin{eqnarray}
    L&\overset{(\ref{geq:npl})(\ref{leq:llzn})}{\leq}&(d-1)L_Z- (d-1)\frac{\rho}{\rho-1} L\\
\Rightarrow&&R_Z=\frac{L}{L_Z}\leq \frac{(\rho-1)(d-1)}{\rho d-1}
\end{eqnarray}

For the second case, i.e., $\rho= +\infty$, from (\ref{geq:nl})(\ref{leq:llzn}), we have
\begin{eqnarray}
    L&\overset{(\ref{geq:nl})(\ref{leq:llzn})}{\leq}&(d-1)L_Z- (d-1)L\\
\Rightarrow&&R_Z=\frac{L}{L_Z}\leq \frac{d-1}{d}
\end{eqnarray}
The proof of the linear converse bound in Theorem \ref{thm:linearnoiserate1} is thus complete.

\section{Proof of Theorem \ref{thm:linearnoiserate2}}\label{pfthm3}
According to Theorem 1 in \cite{Li_Sun_CDS}, achieving the highest rate requires $N = L$ and $\rho = +\infty$. The proof of Theorem \ref{thm:linearnoiserate2} closely follows the analysis of the CDS instance depicted in Fig.~\ref{fig4} in the previous section.

Consider any CDS instance $G_f(V,E)$, and focus on an internal qualified edge $e = \{v_{i_1}, v_{i_2}\}$ within a residing unqualified path $P = \{v_{i_1}, v_{j_1}, \cdots, v_{j_{d-1}}, v_{i_2}\} \subset V$ consisting of $d+1$ nodes and $d$ unqualified edges. Such $e$ and $P$ are guaranteed to exist because the noise rate capacity of 1 condition is violated, ensuring the presence of at least one internal qualified edge and a corresponding residing unqualified path. Without loss of generality, assume these $d+1$ nodes belong to $Q$ qualified components, where $Q \in \{2, \cdots, d\}$\footnote{$Q$ cannot be $1$, because $Q=1$ would imply that all nodes in the residing unqualified path belong to the same qualified component, which in turn means that $\rho < +\infty$.}. Within the $q^{\text{th}}$ qualified component ($q \in \{1, 2, \cdots, Q\}$), suppose there are $d_q$ nodes $v^{(q)}_1, v^{(q)}_2, \cdots, v^{(q)}_{d_q}$ connected to the residing unqualified path.

For any $q$ such that $d_q = 1$, it follows from the constraint $N = L$ and Eq.~(\ref{eq:rv}) that the following condition holds:
\begin{eqnarray}
    r_{v^{(q)}_1}=N=L. \label{eq:nl111}
\end{eqnarray}

For any $q$ such that $d_q \geq 2$, assume the $q^{\text{th}}$ qualified component contains $D_q$ nodes, denoted by $\{v^{(q)}_1, v^{(q)}_2, \cdots,$ $ v^{(q)}_{D_q}\}$, and $\{v^{(q)}_1, v^{(q)}_2, \cdots, v^{(q)}_{d_q}\} \subseteq \{v^{(q)}_1, v^{(q)}_2, \cdots, v^{(q)}_{D_q}\}$. Without loss of generality, assume the qualified edges are $\{v^{(q)}_{i_1}, v^{(q)}_{i_2}\}, \{v^{(q)}_{i_2}, v^{(q)}_{i_3}\}, \cdots, \{v^{(q)}_{i_{D_q-1}}, v^{(q)}_{i_{D_q}}\}$. 
By the property of submodularity, we derive the following inequality:
\begin{eqnarray}
    \alpha_{v^{(q)}_1v^{(q)}_2\cdots v^{(q)}_{d_q}} &\geq& \alpha_{v^{(q)}_1v^{(q)}_2\cdots v^{(q)}_{D_q}} \label{eq:bmm}\\
    &\overset{(\ref{eq:v1v2v3vp})}{\geq}& \alpha_{v^{(q)}_{i_1} v^{(q)}_{i_2}} + \alpha_{v^{(q)}_{i_2} v^{(q)}_{i_3}} + \cdots + \alpha_{v^{(q)}_{i_{D_q-1}}v^{(q)}_{i_{D_q}}} - (D_q-2)N \label{eq:bmm1}\\
    &\overset{(\ref{eq:noise})}{\geq}& (D_q-1)L - (D_q-2)N \label{eq:bmm2}\\
    &\overset{(\ref{eq:nl111})}{=}& L.\label{eq:bmm4}
\end{eqnarray}
where (\ref{eq:bmm}) holds because $\{v^{(q)}_1, v^{(q)}_2, \cdots, v^{(q)}_{d_q}\} \subseteq \{v^{(q)}_1, v^{(q)}_2, \cdots, v^{(q)}_{D_q}\}$. (\ref{eq:bmm1}) holds because we apply submodularity in (\ref{eq:v1v2v3vp}).
Similarly, (\ref{eq:bmm2}) holds because $\{v^{(q)}_{i_1}, v^{(q)}_{i_2}\}, \{v^{(q)}_{i_2}, v^{(q)}_{i_3}\}, \cdots, \{v^{(q)}_{i_{D_q-1}}, v^{(q)}_{i_{D_q}}\}$ are qualified edges, satisfying the conditions for the noise constraint.

Without loss of generality, assume the $1^{th}$ component contain the internal qualified edge $e=\{v_{i_1},v_{i_2}\}$, replace $\alpha_{v^{(q)}_1v^{(q)}_2}$ with $\alpha_{v_{i_1}v_{i_2}}$ in (\ref{eq:bmm1}), we have
\begin{eqnarray}
    \alpha_{v^{(1)}_1v^{(1)}_2\cdots v^{(1)}_{d_1}}&\overset{(\ref{eq:bmm1})}{\geq}&\alpha_{v_{i_1}v_{i_2}}+(D_q-2)L-(D_q-2)N\label{geq:comp}\\
    &\overset{(\ref{eq:nl111})}{=}&\alpha_{v_{i_1}v_{i_2}}\label{eq:bmm5}
\end{eqnarray}
Finally, consider the noise space overlap of $d+1$ nodes: $v_{i_1}, v_{j_1}, \cdots, v_{j_{d-1}}, v_{i_2}$. These nodes are connected through a residing unqualified path that connects to $Q$ components. Assume there are $Q_1$ components with $d_q=1$ and $Q_2$ components with $d_q \ge 2$. 
These numbers satisfy $Q_1+Q_2=Q$.
To connect these $Q$ qualified components, a minimum of $Q-1$ unqualified edges in the residing unqualified path is required. Without loss of generality, assume the $Q-1$ unqualified edges are $\{v^{(1)}_{i_1}, v^{(2)}_{i_2}\}, \{v^{(2)}_{i_3}, v^{(3)}_{i_4}\}, \cdots, \{v^{(Q-1)}_{i_{2Q-3}}, v^{(Q)}_{i_{2Q-2}}\}$. By the property of submodularity, we have:
\begin{eqnarray}
    \alpha_{v_{i_1}v_{j_1}\cdots v_{j_{d-1}}v_{i_2}}&\geq& \sum_{q\in [Q_2]}\alpha_{v^{(q)}_1v^{(q)}_2\cdots v^{(q)}_{d_q}}+\sum_{q\in [Q-1]}\alpha_{v^{(q)}_{i_{2q-1}}v^{(q)}_{i_{2q}}}-Q_2N-(Q-2)N\label{eq:qq1}\\
    &\overset{(\ref{eq:bmm5})}{\geq}&\alpha_{v_{i_1}v_{i_2}}+ \sum_{q\in [Q_2]\setminus\{1\}}\alpha_{v^{(q)}_1v^{(q)}_2\cdots v^{(q)}_{d_q}}+\sum_{q\in [Q-1]}\alpha_{v^{(q)}_{i_{2q-1}}v^{(q)}_{i_{2q}}}-Q_2N-(Q-2)N\label{eq:qq2}\\
    &\overset{(\ref{eq:bmm4})(\ref{geq:lzs})}{\geq}&\alpha_{v_{i_1}v_{i_2}}+(Q_2-1)L+\sum_{q\in [Q-1]}(r_{v^{(q)}_{i_{2q-1}}}+r_{v^{(q)}_{i_{2q}}}-L_Z)-Q_2N-(Q-2)N~~\label{eq:qq3}\\
    &\overset{(\ref{eq:nl111})}{=}&\alpha_{v_{i_1}v_{i_2}}+(Q_2-1)L+(Q-1)(2L-L_Z)-Q_2L-(Q-2)L\\
    &=&\alpha_{v_{i_1}v_{i_2}}+(Q-1)(L-L_Z).\label{eq:qllz}
\end{eqnarray}
where (\ref{eq:qq1}) holds because, by submodularity, the $Q_2$ components connected by $Q_2$ unqualified edges require subtracting $Q_2 N$ symbols, and for $Q-1$ unqualified edges, submodularity requires subtracting $(Q-2) N$ symbols. 
(\ref{eq:qq2}) holds because the first component contains the internal qualified edge $\{v_{i_1}v_{i_2}\}$, to which we apply (\ref{eq:bmm5}).
In (\ref{eq:qq3}), the second term holds since $v^{(q)}_1, v^{(q)}_2, \ldots, v^{(q)}_{d_q}$ belong to the same qualified component, and thus we apply (\ref{eq:bmm4}); the third term holds because $\{v^{(q)}_{i_{2q-1}}v^{(q)}_{i_{2q}}\}$ is an unqualified edge, and we apply (\ref{geq:lzs}).

Then from the correctness constraint (\ref{eq:dec}) for qualified edge $e = \{v_{i_1}, v_{i_2}\}$, we have
\begin{eqnarray}
L &\overset{(\ref{eq:dec})}{\leq}&\mbox{rank}\left({\bf P}_{v_{i_1}} {\bf F}_{v_{i_1}} - {\bf P}_{v_{i_2}} {\bf F}_{v_{i_2}}\right) \\
&\overset{(\ref{eq:st1}) (\ref{eq:st2})}{=}& \mbox{rank} \left( {\bf P}_{v_{i_1}} (\alpha_{v_{i_1}v_{j_1}v_{j_2}\cdots v_{j_{d-1}}v_{i_{2}}} +1:\alpha_{v_{i_1}v_{i_2}}, :) {\bf F}_{v_{i_1}}\right. \notag\\
&&-\left. {\bf P}_{v_{i_2}}(\alpha_{v_{i_1}v_{j_1}v_{j_2}\cdots v_{j_{d-1}}v_{i_{2}}} +1:\alpha_{v_{i_1}v_{i_2}}, :) {\bf F}_{v_{i_2}} \right) \\
&\leq& \alpha_{v_{i_1}v_{i_2}} - \alpha_{v_{i_1}v_{j_1}v_{j_2}\cdots v_{j_{d-1}}v_{i_{2}}} \\
&\overset{(\ref{eq:qllz})}{\leq}&\alpha_{v_{i_1}v_{i_2}}-(\alpha_{v_{i_1}v_{i_2}}+(Q-1)(L-L_Z))\\
&=&(Q-1)L_Z-(Q-1)L\\
\Rightarrow&&R_Z=\frac{L}{L_Z}\leq \frac{Q-1}{Q}
\end{eqnarray}
The proof of the linear converse bound in Theorem \ref{thm:linearnoiserate2} is thus complete.

\section{Proof of Theorem \ref{thm:ach}}\label{pfthm4}
The converse proof is shown in section \ref{pfthm2}.
In this section, we demonstrate that the linear noise rate $(d-1)/d$ is achievable under the condition specified in Theorem \ref{thm:ach}. Specifically, we set $L_Z = d$, meaning each noise consists of $L_Z$ symbols $Z = (Z_0, Z_1, \cdots, Z_{d-1})$ from the finite field $\mathbb{F}_p$. Similarly, we define $L = d-1$, so that each secret comprises $L$ symbols $S = (S_1, \cdots, S_{d-1})$ from $\mathbb{F}_p$. Here, $p$ is assumed to be a prime number no smaller than $2d - 2$.

To prepare for the achievable scheme, we first define $L = d$ generic linear combinations $l_0, l_1, \cdots, l_{d-1}$ of the secret symbols as follows:
\begin{eqnarray}
(l_0; l_1; \cdots; l_{d-1})_{d \times 1} &=& {\bf C}_{d \times (d-1)} \times (S_1; \cdots; S_{d-1})_{(d-1) \times 1}, \notag\\
{\bf C}_{d \times (d-1)}(i, j) &=& \frac{1}{x_i - y_j}, \quad i \in \{1,\cdots, d\}, \, j \in \{1,\cdots, d-1\}, \label{eq:cc}
\end{eqnarray}
where $x_i$ and $y_j$ are distinct elements from $\mathbb{F}_p$. The existence of such elements is guaranteed because the field size $p$ is no smaller than $2d-2$. 
Here, ${\bf C}_{d \times (d-1)}$ is a Cauchy matrix, known for its property that every square sub-matrix has full rank \cite{Schechter}. This ensures that the linear combinations $l_0, l_1, \cdots, l_{d-1}$ are independent and well-defined, which is critical for the achievable scheme.

Consider any CDS instance $G_f(V, E)$ that satisfies the condition in Theorem~\ref{thm:ach}. The graph contains $2(kd+1)$ nodes, $V = \{v_1, v_2, \ldots, v_{2(kd+1)}\}$, forming an unqualified path with edges $\{v_1, v_2\}, \{v_2, v_3\}, \ldots,$ $ \{v_{2kd+1}, v_{2(kd+1)}\}$. The first node is $v_1$ and the last node is $v_{2(kd+1)}$, and the unqualified path distance is $d$. In addition, the graph contains $kd+1$ cyclic qualified edges: $\{v_1, v_{d+1}\}, \{v_3, v_{d+3}\}, \ldots, \{v_{(2k-1)d+2}, v_{2(kd+1)}\},$ $ \{v_{(2k-1)d+4}, v_2\}, \ldots,$ $ \{v_{2kd+1}, v_{d-1}\}$. Here, $k \in \{1, 2, \ldots\}$ and $d \in \{3, 5, \ldots\}$ determine the structure of the graph. With this setup, we can now specify the code assignment.

\subsubsection{Assigning Noise Variables}
We begin by assigning noise variables to each node. For any node $v_i$ where $i \in [(2k-1)d+1]$, the noise variables $v^{(z)}_i$\footnote{We use $v^{(z)}_i$ denote the noise part of node $V_i$.} are assigned circularly as follows:
\begin{eqnarray}
v^{(z)}_i = \{Z_k\}_{k \in [d]_0 \setminus \{i ~\mbox{mod}~ d\}}, \quad i \in [(2k-1)d+1].
\end{eqnarray}
where $[d]_0\triangleq \{0,1,2,\cdots,d-1\}$.

For the nodes $v_{(2k-1)d+2}$ and $v_{2kd+2}$, the noise variables are assigned as:
\begin{eqnarray}
v^{(z)}_{(2k-1)d+2} = v^{(z)}_{2kd+2} = \{Z_0+Z_1, Z_2+Z_1, Z_3+Z_1, \cdots, Z_{d-1}+Z_1\}.
\end{eqnarray}

For the remaining $d-1$ nodes, denoted as $\{v_j \,|\, j \in \{(2k-1)d+3, \cdots, 2kd+1\}\}$, the noise variables are assigned to satisfy the correctness constraint. Specifically, each node is assigned the same noise as its counterpart in the corresponding qualified edge:
\begin{eqnarray}
&&v^{(z)}_j = v^{(z)}_{j-d}, \quad j \in \{(2k-1)d+3, \cdots, 2kd+1\},  j~ \mbox{mod}~ 2 = 0, \label{eq:mod0}\\
&&v^{(z)}_j = v^{(z)}_{j+d-(2kd+2)}, \quad j \in \{(2k-1)d+3, \cdots, 2kd+1\},j~ \mbox{mod}~ 2 = 1. \label{eq:mod1}
\end{eqnarray}

\subsubsection{Assignment of Secrets and Coefficients}
We now describe the assignment of secrets and coefficients. For any node $v_i$ where $i \in [(2k-1)d+1]$, we assign the secret symbols $l_0, S_1, \cdots, S_{d-1}$ to the corresponding noise symbols $Z_0, Z_1, \cdots, Z_{d-1}$, respectively. 
Next, we outline the process of assigning coefficients. Consider the nodes that contain each noise symbol $Z_0, Z_1, \cdots, Z_{d-1}$ sequentially, as well as the unqualified edges associated with each noise symbol $Z_i$, where $i \in \{0, 1, \cdots, d-1\}$. Note that the noise symbols $Z_0, Z_1, \cdots, Z_{d-1}$ are assigned circularly across $d$ noise symbols. 
For each noise $Z_i$, consider the unqualified paths containing it. Each unqualified path consists of at most $d$ nodes and does not include any internal qualified edges. Within a single unqualified path, we assign the same coefficient (i.e., the same message) to ensure security. 
Across different unqualified paths, however, we assign different coefficients (i.e., distinct messages) to maintain independence between paths.
Finally, for the noise $Z_i$ located in the $j$-th unqualified path, the message assignment follows the specific rule outlined below:
\begin{eqnarray}
    &&\mbox{Replace}~ Z_i ~\mbox{with}~ j \times l_{i} + Z_i, ~~\mbox{if}~ i = 0, \label{sch:ieqd1}\\
    &&\mbox{Replace}~ Z_i ~\mbox{with}~ j \times S_{i} + Z_i, ~~\mbox{if}~ i \in [d-1]. \label{sch:ieqd2}
\end{eqnarray}

Suppose the message assigned to node $v_{(2k-1)d+1}$ is as follows:
\begin{eqnarray}
    v_{(2k-1)d+1} = \{J_{0} \times l_{0} + Z_{0}, J_2 \times S_{2} + Z_2, \cdots, J_{d-1} \times S_{d-1} + Z_{d-1}\}, \label{eq:2k1d1}
\end{eqnarray}
where $J_i$ represents the coefficient assigned to each symbol.

To satisfy the security constraint, we introduce an additional noise term $Z_1$ to each symbol in $v_{(2k-1)d+1}$. This ensures that the resulting messages are indistinguishable from random noise to any unauthorized observer. The updated message assigned to node $v_{(2k-1)d+2}$ becomes:
\begin{eqnarray}
    v_{(2k-1)d+2} = \{J_{0} \times l_{0} + Z_{0} + Z_1, J_2 \times S_{2} + Z_2 + Z_1, \cdots, J_{d-1} \times S_{d-1} + Z_{d-1} + Z_1\}. \label{eq:2k1d2}
\end{eqnarray}
By adding $Z_1$, we effectively obfuscate the original message, making it secure while maintaining the structure necessary for correctness.

Next, consider the message assignment for node $v_{(2k-1)d+3}$. Note that $v^{(z)}_{(2k-1)d+3}=v^{(z)}_{(2k-2)d+3}$.
The secret $J_1 \times l_1$ is first assigned to the noise variable $Z_1$. To satisfy the security constraint, we assign the secret $J_{0} \times l_{0} + Z_{0}+Z_1-(J_1\times l_1+Z_1)-Z_0=J_{0} \times l_{0} - J_1\times l_1$ to the noise $Z_0$. 
For any other noise $Z_i$ where $i \in [d-1] \setminus \{0,1,3\}$, we assign the secret $J_i\times S_i +Z_i+Z_1-(J_1\times l_1+Z_1)-Z_i=J_i\times S_i-J_1\times l_1$.

Thus, the messages assigned to $v_{(2k-1)d+3}$ are given by:
\begin{eqnarray}
    && \mbox{Replace}~ Z_i ~\mbox{by}~ J_{0} \times l_{0} - J_1 \times l_1 + Z_i ~~\mbox{if}~ i = 0, \label{sch:ieqd3} \\
    && \mbox{Replace}~ Z_i ~\mbox{by}~  J_1 \times l_1 + Z_i ~~\mbox{if}~ i = 1,  \\
    && \mbox{Replace}~ Z_i ~\mbox{by}~ J_i \times S_i - J_1 \times l_1 + Z_i ~~\mbox{if}~ i \in [d-1] \setminus \{0,1,3\}. \label{sch:ieqd4}
\end{eqnarray}

Consider the $d-2$ nodes $v_{(2k-1)d+4}, \cdots, v_{2kd+1}$ and the assignment of messages to their corresponding noise symbols $Z_0, Z_1, \cdots, Z_{d-1}$. Note that the noise assignment refers to (\ref{eq:mod0})(\ref{eq:mod1}).
For each noise symbol $Z_i$ where $i \in \{0, 1, \cdots, d-1\}$, we process one symbol at a time, and determine its assignment as follows:

\emph{1)} If the noise symbol $Z_i$ is on the same unqualified path as noise symbol in the node $v_{(2k-1)d+3}$,
   \begin{eqnarray}
       &&\text{Assign the same message as in node}~v_{(2k-1)d+3}. \label{sch:d4kd1}
   \end{eqnarray}

\emph{2)} Otherwise (for noise symbols on different unqualified paths),
   \begin{eqnarray}
       &&\text{Replace}~ Z_i ~\text{by}~ J_i \times l_i + Z_i, \quad i \in \{0, 1, \cdots, d-1\}. \label{sch:d4kd2}
   \end{eqnarray}

This strategy ensures that noise symbols within the same unqualified path maintain consistency in their assigned messages, while those in different paths are assigned distinct messages based on their respective coefficients \(J_i\) and secrets \(l_i\). By doing so, the scheme preserves both the correctness and security constraints across all nodes in the unqualified path.

Consider the last node $v_{2(kd+1)}$. To satisfy the security constraint, each symbol associated with the noise terms $Z_i+Z_1$, where $i \in \{0, 2, 3, \cdots, d-1\}$, is assigned as a linear combination of the symbols containing $Z_i$ and $Z_1$ in node $v_{2kd+1}$. The assigned messages for node $v_{2(kd+1)}$ are given by:
\begin{eqnarray}
    v_{2(kd+1)}&=&(J_0 \times l_0 + J_1 \times l_1 + Z_0 + Z_1, J_2 \times l_2 + J_1 \times l_1 + Z_2 + Z_1, \cdots, \notag\\
    &&J_{d-1} \times l_{d-1} + J_1 \times l_1 + Z_{d-1} + Z_1). \label{eq:2kd1}
\end{eqnarray}

This assignment ensures that the noise symbols $Z_i$ and $Z_1$ are appropriately combined to preserve security while maintaining the integrity of the overall scheme.

Now that the code assignment is complete, we move on to verify that the proposed scheme satisfies both correctness and security.

First, we prove that the security constraint (\ref{sec}) is satisfied. 

\emph{Case 1:} Unqualified edges involving $v_{(2k-1)d+2}$ or $v_{2(kd+1)}$. Consider the unqualified edges $\{v_{(2k-1)d+1},$ $v_{(2k-1)d+2}\}$, $\{v_{(2k-1)d+2},v_{(2k-1)d+3}\}$, and $\{v_{2kd+1},v_{2(kd+1)}\}$. According to the code assignments in (\ref{eq:2k1d1}), (\ref{eq:2k1d2}), (\ref{sch:ieqd3}), (\ref{sch:ieqd4}), and (\ref{eq:2kd1}), the message associated with any shared noise between two nodes in these unqualified edges is a linear combination of the corresponding message in the other node. Consequently, no information about the secrets can be inferred, ensuring that (\ref{sec}) holds.

\emph{Case 2:} Other unqualified edges.
For all other unqualified edges, which do not involve $v_{(2k-1)d+2}$ or $v_{2(kd+1)}$, the message assignment rules in (\ref{sch:ieqd1}), (\ref{sch:ieqd2}), (\ref{sch:ieqd3}), (\ref{sch:ieqd4}), (\ref{sch:d4kd1}), and (\ref{sch:d4kd2}) ensure that if the noise is the same unqualified path, the message remains identical. As a result, no additional information about the secrets is revealed.

In conclusion, the security constraint (\ref{sec}) is satisfied for all unqualified edges, guaranteeing the scheme's security.

Second, we prove that the correctness constraint (\ref{dec}) is satisfied. 

\emph{Case 1:} Qualified edges within $v_1,\cdots,v_{(2k-1)d+1}$.
According to the noise assignment, for any qualified edge, the noise symbols $Z_i$ are assigned identically across the nodes. Based on the message assignment rules (\ref{sch:ieqd1}) and (\ref{sch:ieqd2}), each noise $Z_i$ (for $i \in \{1,\cdots,d-1\}$) or $Z_0$ is associated with a distinct secret symbol $S_i$ or $l_0$, respectively. 

Additionally, for a given qualified edge, each shared secret symbol $S_i$ or $l_0$ is multiplied by a unique coefficient $j$, as detailed in (\ref{eq:2k1d1}) and (\ref{eq:2k1d2}). This is because the noise symbols are assigned circularly, and each unqualified path connecting $Z_i$ involves at most $d-1$ nodes. As a result, the coefficients corresponding to the $d-1$ secret symbols in the qualified edge are distinct. Consequently, from any $d-1$ of the $d$ secret symbols $\{S_1,\cdots,S_{d-1},l_0\}$, the secret vector $S=(S_1,\cdots,S_{d-1})$ can be recovered without error.

\emph{Case 2:} Qualified edges involving $v_{(2k-1)d+2},\cdots,v_{2(kd+1)}$.
For any qualified edge that includes at least one node from $v_{(2k-1)d+2}$ to $v_{2(kd+1)}$, the message assignment ensures that each of the $d-1$ noise symbols is mixed with a distinct linear combination of $S_i$ and $l_j$. The coefficients $J_i$ associated with $S_i$ and $l_j$ in these linear combinations are distinct. 

Since the values $l_j$ are derived from a Cauchy matrix (\ref{eq:cc}), every square sub-matrix of this matrix has full rank. This guarantees that from the qualified edge $\{v, u\}$, we can derive $L$ independent linear equations involving $S_i$ and $l_j$. Thus, the secret vector $S$ can be recovered without error.

In conclusion, the correctness constraint (\ref{dec}) is satisfied, ensuring the scheme functions as intended.
The proof of theorem \ref{thm:ach} is now complete.

\section{Conclusion} \label{conclus}
In this work, we studied the noise capacity of conditional disclosure of secrets (CDS) and offered important insights into understanding its structural properties and theoretical limits.
We first established necessary and sufficient conditions which guarantee the  achievability of the extremal case where  the noise capacity reaches it maximum value of one. 
\Ip, it was shown that the unit capacity is achievable if and only if
the CDS graph contains no internal qualified edges within unqualified paths.
This graph-theoretic representation serves as a systematic framework to optimize noise utilization in CDS.
Second, beyond the the above extremal case, we derived a converse (upper) bound on the linear noise rate, which incorporates the  characteristics of the CDS graph, revealing the pivotal impact of the covering parameter and the unqualified path distance on noise efficiency.
Third,  when an additional constraint that the message size should  match the secret size is enforced, we refined the proposed converse bound based on a careful inspection of the qualified components and their interconnections in the CDS graph. 
Finally, we demonstrated  the \achvblty of the proposed converse bounds  through  a specific CDS instance featured by cyclic qualified edges and one unqualified path.

Our work unified the analysis of noise capacity in CDS by addressing its maximum potential, general converse bounds, and achievable instances. The graph-theoretic representation of CDS provides a novel framework for analyzing and designing new CDS schemes.
Nevertheless, the general   optimality of CDS remains open. An immediate direction is to determine whether the proposed linear noise bound is achievable.

\bibliographystyle{IEEEtran}
\bibliography{Thesis}
\end{document}